\documentclass{article}
\usepackage{ijcai26}

\usepackage{times}
\usepackage{soul}
\usepackage{url}
\usepackage[hidelinks]{hyperref}
\usepackage[utf8]{inputenc}
\usepackage[small]{caption}
\usepackage{graphicx}
\usepackage{amsmath}
\usepackage{amsthm}
\usepackage{booktabs}
\usepackage{algorithm}
\usepackage{algorithmic}
\usepackage[switch]{lineno}

\usepackage{amsfonts}  
\usepackage{makecell}
\usepackage{subcaption}
\usepackage{amssymb}
\usepackage{mathtools}
\usepackage{array}
\usepackage{longtable}
\usepackage{enumitem}
\usepackage{xcolor}
\definecolor{darkgreen}{rgb}{0.0,0.5,0.0}
\usepackage{ragged2e}
\usepackage{placeins}
\usepackage{multirow}
\usepackage{siunitx}

\makeatletter
\renewcommand{\p@subfigure}{\thefigure}
\makeatother

\title{SentAttack: A Sentence-Level Black-Box Adversarial Attack Method for Dense Retrieval Models}

\author{
Luping Wei\and
Yamin Hu\and
Sihan Shang\and
Shiyin Wang\and
Wenjian Luo\thanks{Corresponding author.}\\
\affiliations
School of Computer Science and Technology, Harbin Institute of Technology, Shenzhen, China\\
\emails
24s151067@stu.hit.edu.cn,
huyamin@hit.edu.cn,
shangsihan@stu.hit.edu.cn,
24s051011@stu.hit.edu.cn,
luowenjian@hit.edu.cn
}
\begin{document}

\maketitle

\begin{abstract}
    Retrieval-Augmented Generation (RAG) systems typically consist of a dense retrieval (DR) model for initial retrieval and a neural ranking model (NRM) for re-ranking.
    Existing robustness studies in RAG mainly focus on NRMs, while adversarial attacks on DR models are mostly limited to word-level perturbations. 
    For low-ranked target documents that are irrelevant to the query, simple word-level attacks are insufficient to mislead DR models into substantially promoting their rankings.
    To solve these problems, we propose SentAttack, a sentence-level black-box adversarial attack method for DR models.
    SentAttack is designed as a two-stage method.
    In the first stage, SentAttack interacts with the black-box RAG system via iterative retrieval to collect ranked documents and ranking information for training a surrogate DR model.
    In the second stage, SentAttack uses the surrogate DR model to encode and cluster documents relevant to the target query, yielding multiple cluster centroids.
    These centroids are concatenated with the target document at the sentence level to form an initial set of adversarial candidates.
    SentAttack then optimizes these candidates using a query- and centroid-guided objective combined with gradient-guided beam search.
    Extensive experiments demonstrate that SentAttack outperforms existing adversarial attacks on DR models, with especially strong performance on low-ranked target documents.
\end{abstract}

\begin{figure}[t]
  \centering
  \hspace{-0.03\columnwidth}
  \includegraphics[width=1.02\linewidth]{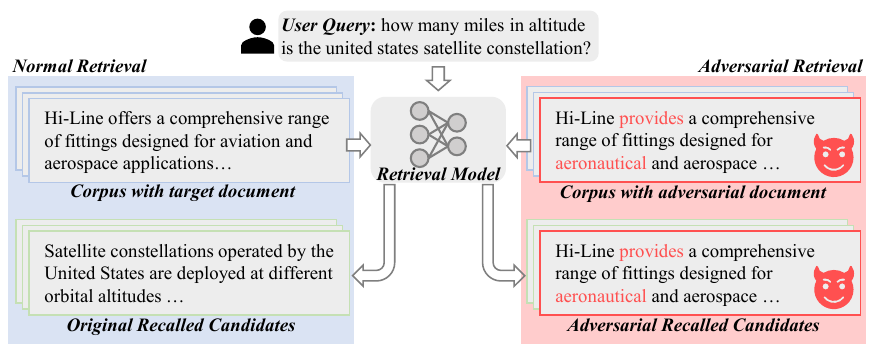}
  \caption{An example of a word-level adversarial attack on a DR model.
  The original target document, ``Hi-Line offers a comprehensive range\ldots'',
  is initially not retrieved for the query. After adversarial perturbation (red text),
  it rises into the top-$K$ candidates, effectively giving Hi-Line advertising.}
  \label{fig:example}
\end{figure}

\section{Introduction}
A typical Retrieval-Augmented Generation (RAG) system consists of two stages: retrieval, which returns an initial top-$K$ documents relevant to the query, and re-ranking, which further re-ranks these candidates \cite{oche2025systematic}.
In the retrieval stage, documents are chunked and embedded into vectors for efficient semantic search using dense retrieval (DR) models, which independently encode queries and documents and achieve high recall in large-scale corpora \cite{guo2022semantic,zhao2024dense}. 
The re-ranking stage typically employs neural ranking models (NRMs), which use interaction-focused architectures to jointly encode queries and documents, producing more precise relevance scores at a higher computational cost than DR models \cite{dai2019deeper,xiong2017end}. 
Although NRMs and DR models substantially improve the retrieval effectiveness of RAG systems, they remain vulnerable to adversarial perturbations, where small textual changes can drastically alter retrieval or ranking outcomes \cite{oche2025systematic}. 
Studying adversarial attacks is therefore crucial for identifying potential vulnerabilities in RAG systems and enhancing system robustness.

Most existing adversarial research in RAG has focused on NRMs, which use interaction-focused architectures to jointly encode queries and documents \cite{liu2022order}.
Comparatively speaking, the adversarial robustness of DR models has received relatively little attention. 
Unlike interaction-centric NRMs, DR models adopt a dual-encoder architecture that independently encodes queries and documents into coarse-grained semantic embeddings \cite{fan2022pretraining,guo2020deep,yates2021bert}.
Consequently, existing NRM attack strategies cannot be directly transferred to DR models \cite{liu2023blackbox}. 
Moreover, current attacks on DR models often fail on low-ranked target documents, which have very low initial relevance to the query and lie far beyond the top-$K$ candidate set. 
This problem is critical for adversarial attacks in RAG systems, since failure in the initial retrieval stage prevents adversarial documents from reaching the re-ranking stage, causing the attack to fail entirely.
Figure~\ref{fig:example} illustrates a simple word-level adversarial attack as an example.

Another problem of existing adversarial attack methods on DR models and NRMs is that they often assume an unrealistically large number of top-$K$ candidates, such as $K=1000$ \cite{liu2023blackbox,wu2023prada}. 
These setups implicitly assume access to an excessively large candidate pool, which is unrealistic in practice.
In reality, DR models typically return only a few top-$K$ documents per query, with $K \in [5, 10]$, due to latency, memory, and LLM input-length constraints \cite{yih2020rag}. 
Consequently, only a few relevant documents reach the generator, limiting the information available for adversarial manipulation.

To solve the above problems, we propose SentAttack, a sentence-level black-box adversarial attack method for DR models.
SentAttack follows a two-stage design.
In the first stage, SentAttack interacts with the target RAG system through iterative retrieval to collect multiple ranked documents for each query, which are then used to construct training data for learning a surrogate dense retrieval model.
In the second stage, iterative retrieval is used to collect query-relevant documents, which are encoded by the surrogate model and clustered to obtain centroid documents. 
These centroids are concatenated with the target document at the sentence level to form an initial set of adversarial candidates, which are then optimized using a query- and centroid-guided objective combined with gradient-guided beam search.
Our main contributions are summarized as follows.
\begin{enumerate}[label=(\arabic*)]
  \item We propose a surrogate DR model training strategy that allows the surrogate DR model to approximate the behavior of a target DR model.
  This strategy obtains positive and negative examples from the top-$K$ documents returned by a black-box RAG system through iterative retrieval, and uses contrastive learning to teach the surrogate DR model to distinguish between these examples.

  \item We propose a sentence-level black-box adversarial attack method that leverages the surrogate DR model to adversarially modify target documents, thereby substantially improving their retrieval rankings.
  
  \item Experimental results on the MS-MARCO Document and MS-MARCO Passage Ranking Datasets demonstrate that our attack method outperforms state-of-the-art baselines, with especially strong performance on low-ranked target documents.
\end{enumerate}

\begin{figure*}[htbp]
  \setlength{\abovecaptionskip}{0pt} 
  
  \centering
  \includegraphics[width=0.90\textwidth]{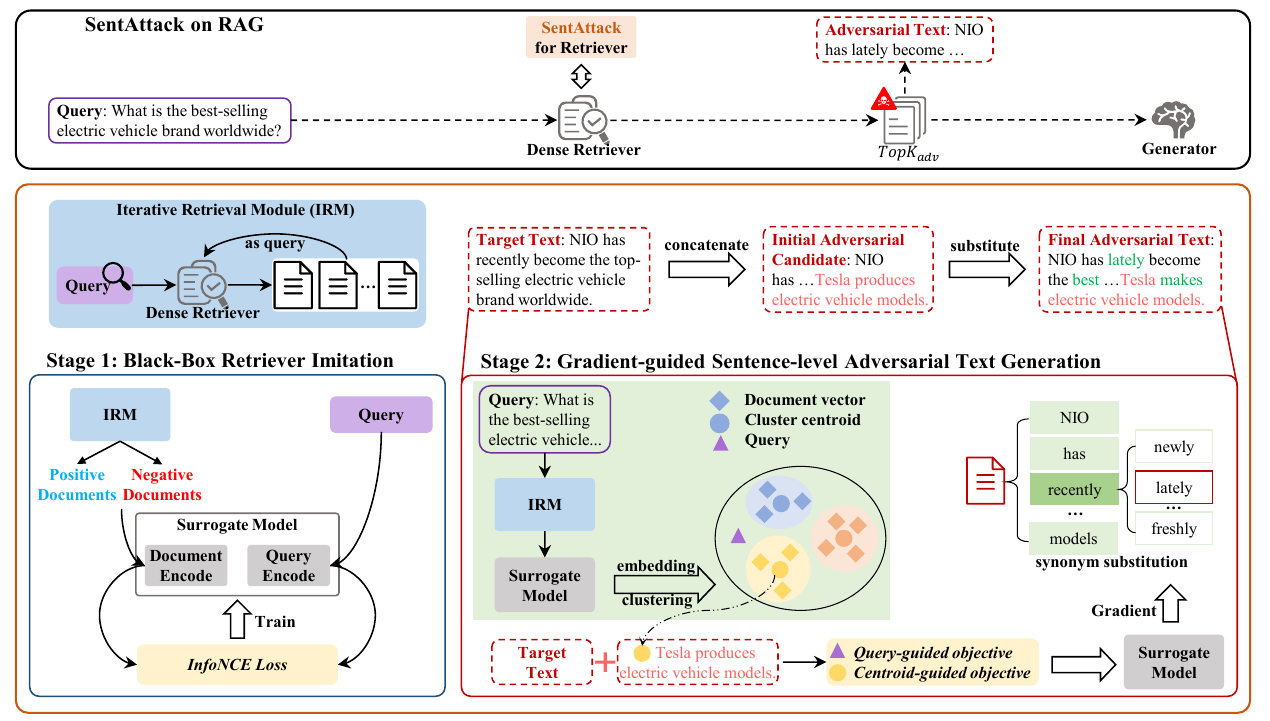}
  
  \caption{Overview of SentAttack.
  For the query ``What is the best-selling electric vehicle brand worldwide?'' (the correct answer: Tesla), the original target document mentioning NIO is not retrieved by the dense retriever. 
  After applying SentAttack to the target document, it becomes highly relevant to the query and is successfully retrieved within the top-$K$ results.
  (NIO is used here only as an illustrative example.)}
  
  \label{fig:overview}
\end{figure*}

\section{Related Work}

\paragraph{Dense Retrieval.}
Existing dense retrieval (DR) models are broadly categorized into single-vector and multi-vector models \cite{lin2022framework}.
Single-vector models encode queries and documents into a single vector for efficient retrieval, but cannot provide fine-grained representations.
Multi-vector models represent documents using multiple vectors to achieve richer semantic representations with the cost of increased storage and computational overhead.
Research on single-vector DR models has primarily focused on improving pre-training objectives \cite{gao2022unsupervised,ma2022contrastive}, refining training procedures \cite{karpukhin2020dpr,khattab2020colbert,zhan2021hardneg}, and performing knowledge distillation \cite{qu2021rocketqa}.
Our work investigates the adversarial robustness of representative single-vector dense retrieval models.

\paragraph{Adversarial Attacks in RAG.}
Existing adversarial attacks in RAG systems primarily target at the re-ranking and generation stages \cite{Liu2024MultiGranular}.
Re-ranking attacks typically rely on word-level substitutions to manipulate the retrieval ranking of target documents, e.g., PRADA \cite{wu2023prada}, which modifies target documents by optimizing a pairwise hinge loss between the document and the initial candidate set.  
Generation-stage attacks often depend on corpus poisoning, e.g., PoisonedRAG \cite{zou2025poisonedrag}, which injects adversarial content into the corpus to influence downstream retrieval and generation.
Our work differs by focusing on attacking DR models in the retrieval stage.
Unlike most retrieval-stage attacks, which assume full access to RAG system components or corpus-specific configurations \cite{liu2023topic},
we consider a black-box setting with limited outputs and obtain query-relevant documents through iterative retrieval to construct training data,
enabling effective attacks without privileged access.

\paragraph{Model Imitation.}
Model imitation is closely related to both model distillation and model extraction in the study of natural language models.
Model distillation typically aims to compress a teacher model by training a student to replicate its predictions, and most distillation methods rely on access to the teacher's logits, which are generally unavailable in black-box settings~\cite{hinton2015distilling}.
By contrast, model imitation (or extraction) does not rely on logits, training a surrogate model using the outputs of the target model~\cite{pal2019framework,krishna2020thieves}. 
Thus, black-box adversarial attacks in RAG often adopt model imitation to recover gradients for optimizing adversarial examples~\cite{wallace2020imitation}. 
Inspired by hard negative sampling~\cite{zhan2021hardneg}, we train a surrogate DR model using contrastive learning on the outputs from the RAG system.

\section{Threat Model}
\label{sec:threat_model}
\paragraph{System Overview.}
A typical RAG system consists of three core components: a dual-encoder retriever, a generator, and a corpus \cite{oche2025systematic}.
In such systems, documents in the corpus are represented as vectors in a shared vector space \cite{oche2025systematic}.
Let the corpus be \( \mathcal{C} = \{ d_1, \dots, d_N \} \), where each document \( d_i \) is encoded offline by the document encoder of the dual-encoder retriever as 
\(
\boldsymbol{e}_{d_i} = E_{{doc}}(d_i) \in \mathbb{R}^m
\).
During querying, a query \( q \) is encoded by the query encoder of the dual-encoder retriever as 
\(
\boldsymbol{e}_q = E_{{qry}}(q)
\),
and the retriever ranks documents using a similarity function \( s(\boldsymbol{e}_q, \boldsymbol{e}_{d_i}) \), 
typically implemented as a dot product or cosine similarity \cite{zhao2024dense}, such that 
\(
s(\boldsymbol{e}_q, \boldsymbol{e}_{d_{(1)}}) \ge s(\boldsymbol{e}_q, \boldsymbol{e}_{d_{(2)}}) \ge \dots \ge s(\boldsymbol{e}_q, \boldsymbol{e}_{d_{(K)}})
\).
The top-\(K\) ranked documents, forming an ordered list 
\(
R = \{ d_{(1)}, \dots, d_{(K)} \}
\), are subsequently fed into the re-ranking stage.

\paragraph{Objective of the Adversary.}
The adversarial attack against retrievers aims to promote a target document \( d_t \) (which is originally ranked outside the top-\(K\)) into the top-\(K\) ordered list \( R \) for a given query \( q \).
This is achieved by finding a perturbation \( p \) such that the resulting adversarial document
\( d_t^{\mathrm{adv}} = d_t \oplus p \)
appears in \( R \).
The perturbation \( p \) is not required to be strictly imperceptible
\cite{Chen2022AdversarialNLP,Li2025CorpusPoisoning}.
Such attacks find applications in real-world scenarios (e.g., search engine advertising \cite{Jafarzadeh2015SEAReview}) and are closely related to corpus poisoning attacks \cite{zou2025poisonedrag}. 
What these scenarios share is that the adversarial or malicious documents are only required to be retrieved and satisfy the adversarial objective, without the need to remain imperceptible.

\paragraph{Capabilities of the Adversary.}
We consider a black-box attack setting in which the adversary has no access to the architecture or training data of the retriever \cite{carlini2021extracting}.
The adversary can input a query \( q \) to the black-box RAG system, and the system returns only a top-$K$ ranked list (typically $K=10$), a restricted level of access consistent with common RAG deployments \cite{yih2020rag}.
In addition, we assume that the adversary can inject adversarial documents into the corpus \cite{zou2025poisonedrag}.
For example, when the corpus is constructed from Wikipedia, the adversary may maliciously edit Wikipedia pages to insert attacker-chosen content.
Prior work has shown that such malicious edits to Wikipedia articles are feasible in practice \cite{carlini-2024-poisoning}.

\paragraph{Target Attack Model.}
Following Liu et al.~\cite{liu2023blackbox}, we perform our experiments on a black-box RAG system.
The RAG system consists of the target retriever and a representative neural ranking model.
The target retriever is implemented using coCondenser~\cite{gao2022unsupervised} and fine-tuned on the corresponding datasets with a two-stage hard-negative sampling strategy \cite{zhan2021hardneg}.
The neural ranking model is realized by PROP~\cite{Ma2021PROP} and fine-tuned using relevance labels together with retrieval results produced by the target retriever \cite{liu2023blackbox}.

\section{Methodology}
\subsection{Overview}
Figure~\ref{fig:overview} illustrates the overall attack method, which aims to improve the ranking of a target document.
The proposed method consists of two stages.

In the first stage, we interact with the target retriever through the Iterative Retrieval Module (IRM)
to obtain ranked documents and sample triplets of positive, hard negative, and random negative examples that capture relative ranking relationships among documents.
Based on these triplets, we construct a training set and train a surrogate retriever
using an InfoNCE-style contrastive loss without relying on human annotations.

In the second stage, we perform sentence-level adversarial optimization on the target document in two steps.
First, IRM is used to retrieve multiple documents relevant to the target query.
These documents are encoded by the surrogate retriever and clustered using a density peak clustering algorithm
to obtain a set of centroid documents.
The target document is then concatenated with the centroid documents to form an initial set of adversarial candidates.
Second, a query- and centroid-guided objective function is used to compute gradients,
which guide synonym substitutions over the initial adversarial candidates, resulting in the final adversarial documents.

\subsection{Black-box Retriever Imitation}
We first prepare a training query set \(Q\). 
For each query \(q \in Q\), we obtain the top-\(K\) documents 
\(\{ d_{1}, \dots, d_{K} \}\) returned by the black-box system.
To expand the set of documents associated with \(q\),
we perform IRM: each document in the top-\(K\) list is treated as a pseudo-query and fed back into the system,
yielding an expanded output of \(K^2\) documents.
We then construct triplets for training the surrogate retriever:
positive examples \(D^{+}\) are sampled from the top-\(K\) list for \(q\),
hard negative examples \(D^{-}_{h}\) are sampled from the expanded IRM outputs,
and random negative examples \(D^{-}_{r}\) are sampled from the retrieval results of other queries in \(Q\).

We train the surrogate retriever \(\tilde{f}\) using contrastive learning on the constructed triplets, 
enabling the surrogate to distinguish between positive, hard negative, and random negative examples \cite{ma2022contrastive}. 
For the surrogate retriever \(\tilde{f}\),
we adopt a vanilla BERT$_{base}$ as the backbone of a dual-encoder retrieval model \cite{Devlin2019BERT}.
The surrogate retriever \(\tilde{f}\) consists of a query encoder \(\tilde{f}_{q}\)
and a document encoder \(\tilde{f}_{d}\),
which encode the query and document into embeddings, respectively:
\begin{equation}
\boldsymbol{e}_q = \tilde{f}_{q}(q), \quad
\boldsymbol{e}_d = \tilde{f}_{d}(d).
\end{equation}
We define the similarity between a query and a document embedding as
\begin{equation}
s(q,d) := \exp\big((\boldsymbol{e}_q)^\top (\boldsymbol{e}_d)/\tau\big),
\end{equation}
where \(\tau\) is a temperature hyperparameter.
For a document set \(D\), the set-level similarity is defined as
\(
s(q, D) = \sum_{d \in D} s(q, d),
\)
which aggregates the similarity between the query and documents in the set.

The InfoNCE-style contrastive loss \cite{ma2022contrastive} over positive, hard negative,
and random negative examples is formulated as
\begin{equation}
\resizebox{\linewidth}{!}{$
\mathcal{L} = - \frac{1}{|Q|} \sum_{q \in Q}
\log \frac{ s(q, D^+) }
         { s(q, D^+)
           + s(q, D^-_{h})
           + s(q, D^-_{r}) }
$}
\end{equation}
This loss encourages the surrogate retriever to capture relative ranking relationships
by contrasting top-\(K\) documents with lower-ranked documents obtained via IRM.

\subsection{Gradient-guided Sentence-level Adversarial Attack}
This stage is performed in two steps:
(1) retrieving documents relevant to the target query using the IRM,
encoding them with the surrogate retriever, clustering them using a density peak clustering algorithm
to obtain centroid documents, and constructing an initial set of adversarial candidates; and
(2) optimizing these candidates using a query- and centroid-guided objective combined with gradient-guided beam search
to produce the final adversarial documents.

\subsubsection{Initial Adversarial Candidates Generation}
To ensure that the retrieved documents for the target query $q_t$ are both relevant and diverse,
we apply the Iterative Retrieval Module. 
We begin by initializing the retrieved document set
\(
R^{(0)} = \{ q_t \},
\)
which contains only the target query itself. Submitting $q_t$ to the black-box system yields
an initial top-$K$ documents: 
\(
R^{(1)} = \{ d^{(1)}_1, \dots, d^{(1)}_K \}.
\)

The iterative retrieval procedure proceeds as follows. 
At each iteration $t = 1, \dots, T-1$, each document in the current set $R^{(t)}$ is treated as a pseudo-query and submitted to the black-box system. 
The retrieved documents from all pseudo-queries are then merged to form the expanded set:
\begin{equation}
  R^{(t+1)} = \cup_{d \in R^{(t)}} \mathcal{R}(d),
\end{equation}  
where $\mathcal{R}(d)$ denotes the set of documents returned by using a single document $d$ as a query. 
This iterative process continues until a sufficient number of documents has been collected.

We denote $\hat{R}$ as the aggregated document set collected across all iterations.
Each document $d \in \hat{R}$ is then encoded by the surrogate retriever as
\(
\boldsymbol{e}_d = \tilde{f}_{d}(d).
\)
We apply the \emph{density peak clustering} algorithm \cite{Chen2020FastDPC} to
\(\{ \boldsymbol{e}_d \}_{d \in \hat{R}}\) to obtain
$n$ document representation centers
\(\{ c_1, \dots, c_n \}\),
each representing a distinct topic related to the query in the embedding space.
Each center $c_i$ is identified as
\begin{equation}
c_i = \boldsymbol{e}_{d_{c_i}}, \quad
\{ d_{c_1}, \dots, d_{c_n} \}
=
\operatorname{Top\text{-}n}_{d \in \hat{R}}
\left( \rho_d \cdot \delta_d \right),
\end{equation}
where $\rho_d$ denotes the local density of \(\boldsymbol{e}_d\) and
$\delta_d$ denotes its minimum distance to any embedding with higher density.
$\operatorname{Top\text{-}n}$ selects the $n$ documents with the largest values of $\rho_d \cdot \delta_d$, 
corresponding to embeddings that are locally dense and far from other higher-density embeddings.

We denote $D_c = \{ d_{c_1}, \dots, d_{c_n} \}$ as the centroid documents corresponding to the representation centers.
To steer the target document $d_t$ toward these representative center directions,
we construct adversarial candidates by concatenating each centroid document $d_{c_i}$ with $d_t$:
\begin{equation}
\tilde{d}_i = d_t \;\oplus\; d_{c_i}, \qquad i = 1, \dots, n,
\end{equation}
where $\oplus$ denotes concatenation.
Each $\tilde{d}_i$ serves as an initial adversarial candidate for gradient-guided refinement.

\subsubsection{Gradient-guided Optimization for Candidates} 
We optimize the initial adversarial candidates by performing gradient-guided synonym substitutions. 
Specifically, we define a query- and centroid-guided attack objective in the embedding space:
\begin{equation}
\mathcal{O}(\tilde{d}_i, q_t, c_i) =
\lambda \, \mathcal{O}_{center}(\tilde{d}_i, c_i)
+ (1 - \lambda) \, \mathcal{O}_{query}(\tilde{d}_i, q_t),
\end{equation}
where, for each $\tilde{d}_i$, the weighting factor $\lambda \in [0,1]$ balances the adversarial candidate's semantic alignment with the centroid document $d_{c_i}$
and its relevance to the target query $q_t$.

\paragraph{Center-guided objective.}
For each initial adversarial candidate $\tilde{d}_i$,
we define a center-guided objective that encourages semantic consistency between the adversarial candidate $\tilde{d}_i$ and the corresponding representation center $c_i$.
The center-guided objective is defined as:
\begin{equation}
\mathcal{O}_{center}(\tilde{d}_i, c_i) =
\left\| \tilde{f}_d(\tilde{d}_i) - c_i \right\|_2^2 .
\end{equation}
This objective mitigates semantic discontinuities caused by document concatenation and encourages alignment between the adversarial candidate and the corresponding centroid document in the embedding space.

\paragraph{Query-guided objective.}
To further bias the adversarial candidate toward the target query $q_t$,
we define a query-guided objective based on the embedding similarity.
The query-guided objective is defined as:
\begin{equation}
\mathcal{O}_{query}(\tilde{d}_i, q_t) =
1 -
\frac{
\tilde{f}_d(\tilde{d}_i)^\top \tilde{f}_q(q_t)
}{
\|\tilde{f}_d(\tilde{d}_i)\| \, \|\tilde{f}_q(q_t)\|
}.
\end{equation}
To promote a higher retrieval ranking of the adversarial candidate for the target query,
this objective encourages increased semantic relevance between the adversarial candidate
and the target query in the embedding space.

\paragraph{Gradient-guided synonym substitution.}
We adopt the projected gradient descent \cite{madry-2018-adversarial} to generate gradient-based adversarial perturbations to the embedding space.
Specifically, for each token $t_i$ in an adversarial candidate $\tilde{d}_i$,
we calculate the gradient $\boldsymbol{g}_{t_i}^{\tilde{f}}$ with respect to the embedding of each token $t_i$ in the surrogate retriever $\tilde{f}$ using the objective $\mathcal{O}$:
\begin{equation}
  \boldsymbol{g}_{t_i}^{\tilde{f}}
  =
  \frac{\partial \mathcal{O}(\tilde{d}_i, q_t, c_i)}
  {\partial \boldsymbol{e}_{t_i}^{\tilde{f}}}.
\end{equation}  
where $\boldsymbol{e}_{t_i}^{\tilde{f}}$ is the embedding of each token $t_i$ obtained by the surrogate model $\tilde{f}$.

Then, the gradient $\ell_2$ norm $I_{t_i} = \|\boldsymbol{g}_{t_i}^{\tilde{f}}\|_2$
reflects the importance of each token $t_i$ for the attack objective \cite{chen2023imperceptible}.
Let $\mathcal{T} = \{t_i^1, t_i^2, \dots, t_i^m\}$ denote the top-$m$ tokens with the highest importance.
For each token $t_i^j \in \mathcal{T}$, we perform substitution by greedily enumerating synonyms from a pre-collected synonym set $\mathcal{C}_{t_i^j}$.
To select a synonym $w \in \mathcal{C}_{t_i^j}$ that better optimizes the attack objective,
we compute the inner product between the embedding $\mathbf{v}_w^{\tilde{f}}$ of the synonym
and the gradient of token $t_i^j$:
\begin{equation}
  s(w, t_i^j)
  =
  (\boldsymbol{g}_{t_i^j}^{\tilde{f}})^\top \mathbf{v}_w^{\tilde{f}}.
\end{equation}
A larger inner product between the token gradient  
and the synonym embedding implies that substituting the synonym 
is expected to decrease the attack objective more effectively.

We employ beam search \cite{liu2022order} to maintain multiple optimal substitution trajectories for each adversarial candidate $\tilde{d}_i$.
At iteration $l$, the beam is defined as
\begin{equation}
\mathcal{B}_i^{(l-1)} = \big\{ \tilde{d}_i^{(l-1,k)} \,\big|\, k = 1, \dots, m \big\},
\end{equation}
where $\mathcal{B}_i^{(l-1)}$ denotes the top-$m$ substitution trajectories retained from iteration $l-1$,
with $\tilde{d}_i^{(l-1,k)}$ representing the $k$-th trajectory in the beam.
Each trajectory in the beam undergoes synonym substitution on its top-$m$ most important tokens, and the $m$ new trajectories with the best objective value are retained to form the updated beam $\mathcal{B}_i^{(l)}$.
The process iterates until a predefined iteration limit is reached or the decrease in the objective value becomes negligible.
The trajectory with the best objective among all explored trajectories is selected as the final adversarial document.

\begin{table*}[htbp]
  \centering
  \resizebox{0.95\textwidth}{!}{
  \begin{tabular}{|l|l|cc|c|cc|c|cc|c|cc|c|}
    \toprule
    & 
    & \multicolumn{3}{c|}{Easy} 
    & \multicolumn{3}{c|}{Middle} 
    & \multicolumn{3}{c|}{Hard} 
    & \multicolumn{3}{c|}{Mixture} \\
    \cmidrule(lr){3-5} \cmidrule(lr){6-8} \cmidrule(lr){9-11} \cmidrule(lr){12-14}
    Dataset & \multicolumn{1}{l|}{Method} 
    & \multicolumn{2}{c|}{SRR} & NRS 
    & \multicolumn{2}{c|}{SRR} & NRS 
    & \multicolumn{2}{c|}{SRR} & NRS 
    & \multicolumn{2}{c|}{SRR} & NRS \\
    \cmidrule(lr){3-14}
    & 
    & @10 & @100 & @100 
    & @10 & @100 & @100 
    & @10 & @100 & @100 
    & @10 & @100 & @100 \\
    \midrule
    \multirow{8}{*}{\makecell[l]{MS-MARCO\\Document}}
    & TF-IDF  & 16.0 & 40.9 & 32.1 & 11.1 & 28.0 & 23.6 & 4.2 & 14.4 & 13.6 & 10.3 & 28.6 & 23.2 \\
    & TS      & 37.8 & 88.1 & 67.5 & 27.2 & 58.0 & 60.3 & 15.1 & 35.8 & 33.5 & 23.0 & 58.0 & 53.8 \\
    & PAT     & 26.5 & 70.2 & 52.2 & 13.7 & 36.0 & 32.0 & 7.9  & 27.1 & 26.4 & 12.7 & 44.5 & 37.8 \\
    & PRADA   & 28.4 & 74.7 & 56.2 & 18.5 & 43.1 & 37.9 & 11.2 & 33.0 & 33.3 & 19.3 & 50.3 & 44.1 \\
    & MCARA   & 43.5 & \underline{92.3} & 73.1 & 28.1 & 66.5 & 61.4 & \underline{24.4} & \underline{50.2} & \underline{51.3} & \underline{32.0} & \underline{72.0} & \underline{62.0} \\
    \cmidrule(lr){2-14}
    & SentAttack$_{Concat}$ & 27.4 & 73.2 & 54.6 & 15.8 & 39.1 & 28.2 & 4.5 & 14.2 & 13.4 & 15.8 & 40.6 & 31.3 \\
    & SentAttack$_{Syn}$    
      & \underline{47.2} & 90.6 & \underline{82.4} 
      & \underline{32.8} & \underline{75.7} & \underline{71.2} 
      & 15.6 & 37.4 & 35.3 
      & 29.8 & 60.3 & 54.3 \\
    & SentAttack                     
      & \textbf{{55.4}} & \textbf{{93.7}} & \textbf{{85.6}} 
      & \textbf{{51.9}} & \textbf{{84.1}} & \textbf{{82.3}} 
      & \textbf{{34.9}} & \textbf{{57.5}} & \textbf{{55.7}} 
      & \textbf{{36.1}} & \textbf{{75.4}} & \textbf{{72.8}} \\
    \midrule
    & & @100 & @1000 & @1000 & @100 & @1000 & @1000 & @100 & @1000 & @1000 & @100 & @1000 & @1000 \\
    \midrule
    \multirow{8}{*}{\makecell[l]{MS-MARCO\\Passage}}
    & TF-IDF  & 10.2 & 35.2 & 25.1 & 6.4  & 19.8 & 18.3 & 2.1  & 10.5 & 10.3 & 6.1  & 21.6 & 17.8 \\
    & TS      & 28.6 & 79.0 & 59.1 & 17.2 & 50.8 & 48.7 & 8.4  & 27.6 & 26.9 & 17.8 & 52.0 & 44.4 \\
    & PAT     & 16.4 & 62.3 & 46.7 & 9.4  & 30.0 & 28.6 & 5.3  & 23.4 & 21.5 & 10.4 & 38.6 & 32.3 \\
    & PRADA   & 28.4 & 68.2 & 51.0 & 13.8 & 39.9 & 39.6 & 10.6 & 31.5 & 30.1 & 14.7 & 46.4 & 40.2 \\
    & MCARA   & 20.1 & 83.1 & 65.9 & 22.7 & 57.3 & 53.7 & \underline{15.3} & \underline{41.1} & \underline{40.2} & 23.7 & \underline{60.5} & \underline{53.3} \\
    \cmidrule(lr){2-14}
    & SentAttack$_{Concat}$ & 18.1 & 65.9 & 53.9 & 10.7 & 37.4 & 31.5 & 2.2 & 12.1 & 11.6 & 9.7 & 37.9 & 28.4 \\
    & SentAttack$_{Syn}$    
      & \underline{42.3} & \underline{84.2} & \underline{70.9} 
      & \underline{38.6} & \underline{76.4} & \underline{72.1} 
      & 12.1 & 32.5 & 30.9 
      & \underline{30.4} & 58.7 & 52.9 \\
    & SentAttack                     
      & \textbf{{50.6}} & \textbf{{92.8}} & \textbf{{83.2}} 
      & \textbf{{45.9}} & \textbf{{83.2}} & \textbf{{80.5}} 
      & \textbf{{31.2}} & \textbf{{54.9}} & \textbf{{53.3}} 
      & \textbf{{33.9}} & \textbf{{65.3}} & \textbf{{63.1}} \\
    \bottomrule
  \end{tabular}}
  \caption{Comparison of adversarial attack methods and SentAttack variants on MS-MARCO Document and Passage Ranking Dataset under different difficulty levels. The best results in each column are highlighted in bold, while the second-best results are underlined.}
  \label{tab:combined_results}
\end{table*}

\section{Experiments}
\subsection{Experimental Settings}
\paragraph{Datasets.}
We evaluate our method on two widely used retrieval benchmarks:
the {\bfseries MS-MARCO Document Ranking Dataset} (hereafter referred to as the {\bfseries Document Dataset}) \cite{nguyen2016msmarco},
and the {\bfseries MS-MARCO Passage Ranking Dataset} (hereafter referred to as the {\bfseries Passage Dataset}) \cite{nguyen2016msmarco}.
Both datasets are collected using Bing search and reflect realistic web retrieval scenarios \cite{nguyen2016msmarco}.
For each dataset, we randomly sample 300 development queries as target queries for evaluation.
For each query, we sample 30 target documents that are initially ranked outside the top-$K$ results of the target retriever,
and categorize them into Easy, Middle, and Hard based on their original relevance score rankings with respect to the query.
We further construct a Mixture set by randomly sampling 10 documents from the 30 targets to ensure diverse attack difficulty.

\paragraph{Implementation details.}
Following prior work \cite{chen2023imperceptible,wu2023prada},
we define three levels of attack difficulty based on the original ranking positions of target documents.
For the Document Dataset, Easy, Middle, and Hard documents are sampled from the rank ranges $[101,200]$, $[201,1000]$, and outside the top $1000$, respectively.
For the Passage Dataset, Easy, Middle, and Hard passages are sampled from the rank ranges $[1000,2000]$, $[2000,10000]$, and outside the top $10000$, respectively.
We set the weighting factor of the attack objective to $\lambda = 0.4$, set the number of cluster centers for constructing adversarial candidates to $n = 5$, and set the synonym substitution constraint threshold to $\rho = 0.8$, which defines the minimum semantic similarity for substitutions and controls the size of the candidate synonym set to ensure fluency, and use a beam search with beam width $m = 3$.
We fix the size of the initial candidate set returned by the target retriever for all datasets (i.e., K=10).

\paragraph{Evaluation metrics.}
Following \cite{liu2023blackbox}, we evaluate the performance of adversarial attacks using two automatic metrics.
(i) \emph{Success Recall Rate (SRR)}@k (\%) measures the proportion of target documents
that are successfully retrieved into the top-$k$ results after adversarial attack.
(ii) \emph{Normalized Ranking Shifts (NRS)}@k (\%) quantifies the relative ranking improvement
of target documents after adversarial attack, conditioned on being successfully recalled into the top-$k$ results, i.e., $\mathrm{NRS}\text{@}k = (\Pi_d - \Pi_{d_{{adv}}}) / \Pi_d \times 100\%$,
where $\Pi_d$ and $\Pi_{d_{{adv}}}$ denote the ranking positions of the original document $d$
and the adversarial document $d_{{adv}}$, respectively.
If $d_{{adv}}$ does not appear in the top-$k$ retrieved results,
its NRS is set to $0$. Note that smaller values of $k$ correspond to a more challenging attack setting
and a stricter evaluation criterion.

\paragraph{Baselines.}
We compare our method with several representative attack approaches as follows.
\begin{itemize}
    \item \textbf{Term Spamming (TS)} \cite{gyongyi2005webspam}: an insertion-based attack method that randomly selects a starting position in the target document and replaces words with terms sampled from the target query.
    \item \textbf{TF-IDF} \cite{ramos-2003-tfidf}: a synonym substitution method that replaces important words in the target document, identified by the highest TF-IDF scores with respect to the target query, with their synonyms.
    \item \textbf{PRADA} \cite{wu2023prada}: a decision-based black-box ranking attack method originally designed for neural ranking models via word substitution, adapted to retrieval models using the pairwise hinge loss between the target document and candidate documents.
    \item \textbf{PAT} \cite{liu2022order}: an anchor-based ranking attack method originally designed for neural ranking models via trigger generation, adapted to retrieval models using the pairwise loss between the target document and the top-1 candidate document.
    \item \textbf{MCARA} \cite{liu2023blackbox}: a contrastive learning-based adversarial attack method on retrieval models that generates perturbations by enforcing consistency between multi-view representations of the target document and their viewers.
\end{itemize}

\subsection{Comparison of Attack Performance}
As shown in Table~\ref{tab:combined_results}, we compare different adversarial attack methods on the Document Dataset and Passage Dataset under varying difficulty levels.
The full SentAttack achieves the strongest attack effectiveness across all settings, with SRR@10 of 55.4/51.9/34.9 on the Document Dataset and SRR@100 of 50.6/45.9/31.2 on the Passage Dataset for Easy/Middle/Hard targets.
MCARA has previously been regarded as one of the strongest adversarial attacks against dense retrievers by leveraging multi-view representations; however, its attack effectiveness drops substantially on Hard targets.
On the Document Dataset, SentAttack improves SRR@10 on Hard targets from 24.4 (MCARA) to 34.9.
On the Passage Dataset, SRR@100 increases from 15.3 to 31.2, more than doubling MCARA’s performance.

We further analyze the contributions of individual components in SentAttack by comparing two ablated variants.
SentAttack$_{Concat}$ only performs centroid document concatenation without applying synonym substitution.
This variant is effective only on Easy targets, while its attack effectiveness drops sharply on Middle and Hard targets.
On the Document Dataset, SRR@10 decreases from 27.4 on Easy targets to 15.8 on Middle and further to 4.5 on Hard targets.
SentAttack$_{Syn}$ removes centroid concatenation and relies solely on gradient-guided synonym substitution.
It achieves strong attack performance on Easy and Middle targets, often ranking as the second most effective method.
However, its effectiveness drops notably on Hard targets, with SRR@10 of 15.6 on the Document Dataset and SRR@100 of 12.1 on the Passage Dataset.
These results indicate that the full SentAttack improves the initial relevance between target documents and queries through centroid concatenation,
providing a stronger starting point for subsequent gradient-guided optimization.

\subsection{Impact of Key Hyperparameters}

\begin{figure}[t] 
  \centering
  \includegraphics[width=1.0\columnwidth]{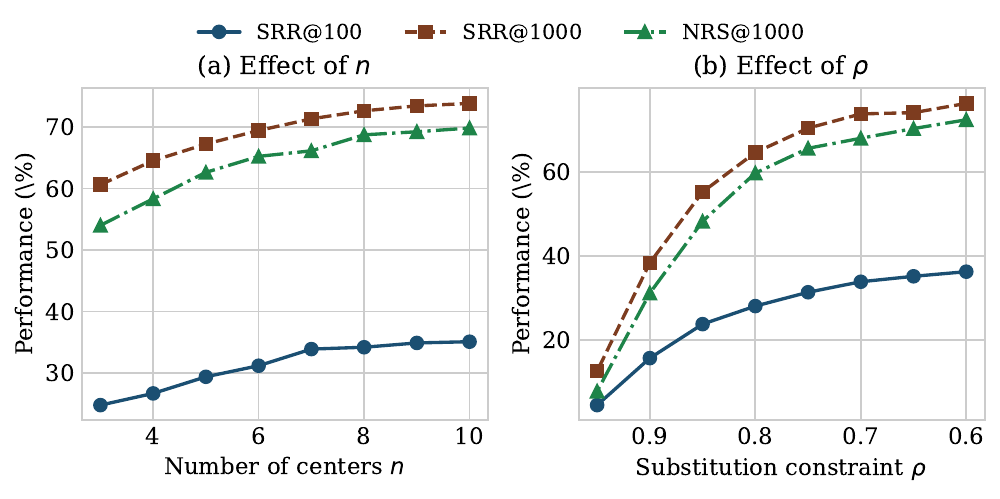}
  \caption{Attack performance of SentAttack under varying hyperparameters on Mixture level in MS-MARCO Passage Ranking Dataset.}
  \label{fig:attack_performance}
\end{figure}

\noindent\textbf{Number of cluster centers.}
We analyze the impact of the number of cluster centers \(n\).
As shown in Figure~\ref{fig:attack_performance} (left), increasing \(n\) improves attack performance by providing more diverse adversarial candidates.
Specifically, SRR@1000 increases from 60.6\% at \(n=3\) to 69.4\% at \(n=6\),
while further increasing \(n\) leads to diminishing improvements, with less than a 4.5\% absolute gain when \(n\) grows from 6 to 10.
NRS@1000 exhibits a similar saturation trend, suggesting that using a large number of cluster centers mainly increases computational cost without clear performance benefits.

\noindent\textbf{Synonym substitution constraint \(\rho\).}
We analyze the impact of the synonym substitution constraint \(\rho\), which defines the minimum semantic similarity threshold to control the size of the candidate synonym set.
As shown in Figure~\ref{fig:attack_performance} (right), relaxing the substitution constraint (i.e., using a smaller \(\rho\)) significantly enhances attack effectiveness by enlarging the pool of candidate synonyms.
For example, SRR@1000 decreases from 64.7\% at \(\rho=0.80\) to 38.4\% at \(\rho=0.90\), with NRS@1000 dropping from 59.8\% to 31.2\%.
However, overly loose constraints may degrade text fluency, revealing a clear trade-off between attack effectiveness and text quality.

\begin{figure}[t]
  \centering
  \includegraphics[width=1.0\columnwidth]{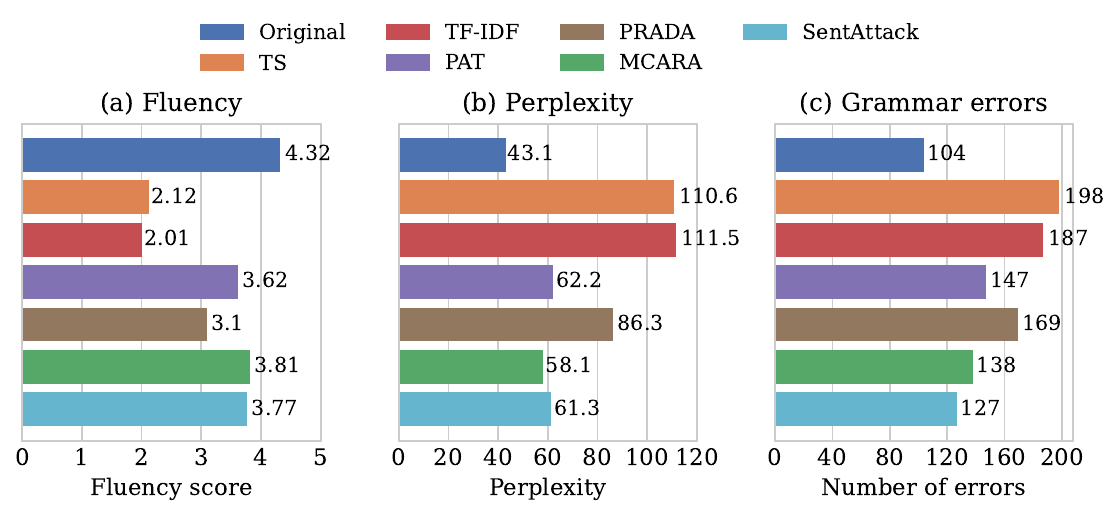}
  \caption{Naturalness comparison of adversarial examples generated by different attack on the MS-MARCO Document Ranking Dataset.}
  \label{fig:naturalness}
\end{figure}

\subsection{Naturalness Evaluation of Adversarial Examples}
We evaluate the naturalness of Mixture-level adversarial examples on the Document Dataset, with similar findings on Passage Dataset.
Following prior work~\cite{liu2022order}, we randomly sample 32 examples per method and assess fluency \cite{holtzman2019curious} , perplexity (PPL), and grammatical correctness.
Fluency is rated by GPT-4o-mini (1--5 scale), PPL is measured using a language model, and grammatical errors are detected by Cheg.\footnote{\url{https://www.chegg.com}} and Gram.\footnote{\url{https://www.grammarly.com}}, aggregated across both tools.
As shown in Figure~\ref{fig:naturalness}, SentAttack achieves comparable fluency to MCARA (3.77 vs.\ 3.81), moderate PPL (61.3 vs.\ 58.1), and fewer grammatical errors (127 vs.\ 138), indicating a favorable balance between naturalness and attack effectiveness.

\subsection{Adversarial Examples Against DR Models during the Re-ranking Stage}
We investigate whether adversarial examples generated by SentAttack against DR models remain highly ranked when processed by the neural re-ranker within the RAG pipeline.
For each dataset, the candidate set returned by the target DR model, including successful adversarial examples, is directly fed into the re-ranker.
The final re-ranked list size is fixed to 100 for MS-MARCO Document and 1000 for MS-MARCO Passage.
Avg.rank denotes the average rank of adversarial examples, while T50\% and T10\% indicate the proportion of adversarial examples appearing in the top 50\% and top 10\% of the final list.
As shown in Table~\ref{tab:attacking_against_nrm}, a noticeable fraction of adversarial examples remains highly ranked.
In particular, up to 13.5\% (Document Dataset) and 12.4\% (Passage Dataset) of adversarial examples in the Easy setting appear in the top 10\%, demonstrating that SentAttack adversarial examples remain effective against the re-ranker.

\begin{table}[!t]
  \centering
  \resizebox{\columnwidth}{!}{
  \begin{tabular}{|l|S[table-format=3.1]|S[table-format=2.1]|S[table-format=2.1]
                   |S[table-format=4.1]|S[table-format=2.1]|S[table-format=2.1]|}
    \toprule
    \multirow{2}{*}{Difficulty} 
    & \multicolumn{3}{c|}{MS-MARCO Document} 
    & \multicolumn{3}{c|}{MS-MARCO Passage} \\
    \cmidrule(lr){2-4} \cmidrule(lr){5-7}
    & {Avg.rank} & {T50\%} & {T10\%} 
    & {Avg.rank} & {T50\%} & {T10\%}  \\
    \midrule
    Easy    & 63.5 & 40.1 & 13.5 & 647.6 & 37.2 & 12.4 \\
    Middle  & 74.8 & 28.7 & 8.6  & 754.1 & 24.6 & 6.9  \\
    Hard    & 89.7 & 13.4 & 4.9  & 860.3 & 10.7 & 2.3  \\
    Mixture & 75.4 & 27.6 & 8.2  & 742.1 & 23.5 & 7.4  \\
    \bottomrule
  \end{tabular}}
  \caption{Effectiveness of SentAttack adversarial examples against neural ranking models (NRMs).}
  \label{tab:attacking_against_nrm}
\end{table}

\subsection{Black-box vs. White-box Attacks}

We compare SentAttack under black-box and white-box settings on Mixture-level targets (Table~\ref{tab:wh_com}).
\begin{table}[H]
  \centering
  \setlength{\tabcolsep}{4pt}
  \resizebox{\columnwidth}{!}{
  \begin{tabular}{|l|S[table-format=2.1]S[table-format=2.1]|S[table-format=2.1]
                  |S[table-format=2.1]S[table-format=2.1]|S[table-format=2.1]|}
    \toprule
    & \multicolumn{3}{c|}{MS-MARCO Document} 
    & \multicolumn{3}{c|}{MS-MARCO Passage} \\
    \cmidrule(lr){2-4} \cmidrule(lr){5-7}
    \multicolumn{1}{|c|}{Method}
    & \multicolumn{2}{c|}{SRR} & NRS
    & \multicolumn{2}{c|}{SRR} & NRS \\
    \cmidrule(lr){2-4} \cmidrule(lr){5-7}
           & {@10} & {@100} & {@100}
           & {@100} & {@1000} & {@1000} \\
    \midrule
    SentAttack               
      & 35.2 & 74.3 & 71.5 
      & 32.7 & 65.3 & 63.2 \\
    SentAttack$_{white}$     
      & 38.8 & 77.1 & 74.2 
      & 34.5 & 68.1 & 65.4 \\
    \bottomrule
  \end{tabular}}
  \caption{Comparison between black-box SentAttack and its white-box variant on Mixture-level targets in the MS-MARCO Document and Passage Ranking Dataset.}
  \label{tab:wh_com}
\end{table}
In the black-box setting, SentAttack trains a surrogate model using the Iterative Retrieval Module (IRM), which iteratively retrieves relevant documents.
In contrast, SentAttack$_{white}$ directly uses the target retriever to generate adversarial examples in a white-box manner.
SentAttack achieves competitive performance: on the Document Dataset,
it attains SRR@10/100 of 35.2/74.3 (vs.\ 38.8/77.1 for SentAttack$_{white}$),
and on the Passage Dataset, SRR@100/1000 of 32.7/65.3 (vs.\ 34.5/68.1).
These results demonstrate that IRM effectively trains a surrogate model even with limited top-K outputs from the target retriever.

\section{Conclusion}
In this paper, we propose SentAttack, a sentence-level black-box adversarial attack method for dense retrieval models in RAG systems.  
Extensive experiments on the MS-MARCO Document and Passage Datasets demonstrate that SentAttack consistently outperforms existing baselines, with particularly strong performance on hard target examples. 
Future work will explore adaptations of SentAttack to multi-turn and multimodal retrieval scenarios, as well as effective defense strategies against adversarial attacks.

\FloatBarrier
\bibliographystyle{named}
\bibliography{densemodel_attack}

@article{guo2022semantic,
  author = {Jiafeng Guo and Yinqiong Cai and Yixing Fan and Fei Sun and Ruqing Zhang and Xueqi Cheng},
  title = {Semantic Models for the First-stage Retrieval: A Comprehensive Review},
  journal = {ACM Transactions on Information Systems},
  volume = {40},
  number = {4},
  pages = {1--42},
  year = {2022},
  publisher = {ACM}
}

@article{zhao2024dense,
  author  = {Wayne Xin Zhao and Jin Liu and Rui Ren and Jun Xu and Ji-Rong Wen},
  title   = {Dense Text Retrieval Based on Pretrained Language Models: A Survey},
  journal = {ACM Transactions on Information Systems},
  volume  = {42},
  number  = {4},
  pages   = {1--60},
  year    = {2024},
  publisher = {ACM}
}

@inproceedings{dai2019deeper,
  author    = {Zhuyun Dai and Jamie Callan},
  title     = {Deeper Text Understanding for IR with Contextual Neural Language Modeling},
  booktitle = {Proceedings of the 42nd International ACM SIGIR Conference on Research and Development in Information Retrieval},
  pages     = {985--988},
  publisher = {ACM},
  year      = {2019}
}

@inproceedings{xiong2017end,
  author    = {Chenyan Xiong and Zhuyun Dai and Jamie Callan and Zhiyuan Liu and Russell Power},
  title     = {End-to-end Neural Ad-hoc Ranking with Kernel Pooling},
  booktitle = {Proceedings of the 40th International ACM SIGIR Conference on Research and Development in Information Retrieval},
  pages     = {55--64},
  publisher = {ACM},
  year      = {2017}
}

@inproceedings{liu2022order,
  author    = {Jiawei Liu and Yangyang Kang and Di Tang and Kaisong Song and Changlong Sun and Xiaofeng Wang and Wei Lu and Xiaozhong Liu},
  title     = {Order-Disorder: Imitation Adversarial Attacks for Black-box Neural Ranking Models},
  booktitle = {Proceedings of the 2022 ACM SIGSAC Conference on Computer and Communications Security (CCS)},
  pages     = {2025--2039},
  publisher = {ACM},
  year      = {2022}
}

@inproceedings{liu2023topic,
  author    = {Yu-An Liu and Ruqing Zhang and Jiafeng Guo and Maarten de Rijke and Wei Chen and Yixing Fan and Xueqi Cheng},
  title     = {Topic-oriented Adversarial Attacks against Black-box Neural Ranking Models},
  booktitle = {Proceedings of the 46th International ACM SIGIR Conference on Research and Development in Information Retrieval},
  publisher = {ACM},
  year      = {2023}
}

@article{wu2023prada,
  author    = {Chen Wu and Ruqing Zhang and Jiafeng Guo and Maarten de Rijke and Yixing Fan and Xueqi Cheng},
  title     = {PRADA: Practical Black-Box Adversarial Attacks against Neural Ranking Models},
  journal   = {ACM Transactions on Information Systems (TOIS)},
  volume    = {41},
  number    = {4},
  pages     = {Article 89},
  year      = {2023}
}

@article{fan2022pretraining,
  author    = {Yixing Fan and Xiaohui Xie and Yinqiong Cai and Jia Chen and Xinyu Ma and Xiangsheng Li and Ruqing Zhang and Jiafeng Guo and others},
  title     = {Pre-training Methods in Information Retrieval},
  journal   = {Foundations and Trends in Information Retrieval},
  volume    = {16},
  number    = {3},
  pages     = {178--317},
  year      = {2022}
}

@article{guo2020deep,
  author    = {Jiafeng Guo and Yixing Fan and Liang Pang and Liu Yang and Qingyao Ai and Hamed Zamani and Chen Wu and W. Bruce Croft and Xueqi Cheng},
  title     = {A Deep Look into Neural Ranking Models for Information Retrieval},
  journal   = {Information Processing \& Management},
  volume    = {57},
  number    = {6},
  pages     = {102067},
  year      = {2020}
}

@inproceedings{yates2021bert,
  author    = {Andrew Yates and Rodrigo Nogueira and Jimmy Lin},
  title     = {Pretrained Transformers for Text Ranking: BERT and Beyond},
  booktitle = {Proceedings of the 14th ACM International Conference on Web Search and Data Mining (WSDM)},
  pages     = {1154--1156},
  publisher = {ACM},
  year      = {2021}
}

@inproceedings{liu2023blackbox,
  author    = {Yu-An Liu and Ruqing Zhang and Jiafeng Guo and others},
  title     = {Black-box Adversarial Attacks against Dense Retrieval Models: A Multi-view Contrastive Learning Method},
  booktitle = {Proceedings of the 32nd ACM International Conference on Information and Knowledge Management (CIKM)},
  pages     = {1647--1656},
  publisher = {ACM},
  year      = {2023}
}

@inproceedings{gyongyi2005webspam,
  author    = {Zoltan Gyongyi and Hector Garcia-Molina},
  title     = {Web Spam Taxonomy},
  booktitle = {Proceedings of the First International Workshop on Adversarial Information Retrieval on the Web (AIRWeb)},
  year      = {2005}
}

@article{lin2022framework,
  author    = {Jimmy Lin},
  title     = {A Proposed Conceptual Framework for a Representational Approach to Information Retrieval},
  journal   = {ACM SIGIR Forum},
  volume    = {55},
  year      = {2022}
}

@inproceedings{karpukhin2020dpr,
  author    = {Vladimir Karpukhin and Barlas Oguz and Sewon Min and Patrick Lewis and Ledell Wu and Sergey Edunov and Danqi Chen and Wen-tau Yih},
  title     = {Dense Passage Retrieval for Open-Domain Question Answering},
  booktitle = {Proceedings of the 2020 Conference on Empirical Methods in Natural Language Processing (EMNLP)},
  publisher = {ACL},
  year      = {2020}
}

@inproceedings{gao2022unsupervised,
  author    = {Luyu Gao and Jamie Callan},
  title     = {Unsupervised Corpus Aware Language Model Pre-training for Dense Passage Retrieval},
  booktitle = {Proceedings of the 60th Annual Meeting of the Association for Computational Linguistics (ACL)},
  pages     = {2843--2853},
  publisher = {ACL},
  year      = {2022}
}

@inproceedings{ma2022contrastive,
  author    = {Xinyu Ma and Ruqing Zhang and Jiafeng Guo and Yixing Fan and Xueqi Cheng},
  title     = {A Contrastive Pre-training Approach to Discriminative Autoencoder for Dense Retrieval},
  booktitle = {Proceedings of the 31st ACM International Conference on Information and Knowledge Management (CIKM)},
  pages     = {4314--4318},
  publisher = {ACM},
  year      = {2022}
}

@inproceedings{khattab2020colbert,
  author    = {Omar Khattab and Matei A. Zaharia},
  title     = {ColBERT: Efficient and Effective Passage Search via Contextualized Late Interaction over BERT},
  booktitle = {Proceedings of the 43rd International ACM SIGIR Conference on Research and Development in Information Retrieval},
  pages     = {39--48},
  year      = {2020}
}

@inproceedings{qu2021rocketqa,
  author    = {Yingqi Qu and Yuchen Ding and Jing Liu and Kai Liu and Ruiyang Ren and Wayne Xin Zhao and Daxiang Dong and Hua Wu and Haifeng Wang},
  title     = {RocketQA: An Optimized Training Approach to Dense Passage Retrieval for Open-Domain Question Answering},
  booktitle = {Proceedings of the 2021 Conference of the North American Chapter of the Association for Computational Linguistics (NAACL)},
  publisher = {ACL},
  year      = {2021}
}

@inproceedings{zhan2021hardneg,
  author    = {Jingtao Zhan and Jiaxin Mao and Yiqun Liu and Jiafeng Guo and Min Zhang and Shaoping Ma},
  title     = {Optimizing Dense Retrieval Model Training with Hard Negatives},
  booktitle = {Proceedings of the 44th International ACM SIGIR Conference on Research and Development in Information Retrieval},
  pages     = {1503--1512},
  year      = {2021}
}

@article{hinton2015distilling,
  author       = {Geoffrey Hinton and Oriol Vinyals and Jeff Dean},
  title        = {Distilling the Knowledge in a Neural Network},
  journal      = {arXiv preprint arXiv:1503.02531},
  year         = {2015},
  url          = {https://arxiv.org/abs/1503.02531}
}

@inproceedings{krishna2020thieves,
  author    = {Kalpesh Krishna and Gaurav Singh Tomar and Ankur P. Parikh and Nicolas Papernot and Mohit Iyyer},
  title     = {Thieves on Sesame Street! Model Extraction of BERT-based APIs},
  booktitle = {Proceedings of the 8th International Conference on Learning Representations (ICLR)},
  year      = {2020},
  address   = {Addis Ababa, Ethiopia},
  url       = {https://openreview.net/forum?id=Byl5NREFDr}
}

@article{pal2019framework,
  author    = {Soham Pal and Yash Gupta and Aditya Shukla and Aditya Kanade and Shirish Shevade and Vinod Ganapathy},
  title     = {A Framework for the Extraction of Deep Neural Networks by Leveraging Public Data},
  journal   = {arXiv preprint arXiv:1905.09165},
  year      = {2019},
  url       = {https://arxiv.org/abs/1905.09165}
}

@inproceedings{wallace2020imitation,
  author    = {Eric Wallace and Mitchell Stern and Dawn Song},
  title     = {Imitation Attacks and Defenses for Black-box Machine Translation Systems},
  booktitle = {Proceedings of the 2020 Conference on Empirical Methods in Natural Language Processing (EMNLP)},
  pages     = {5531--5546},
  publisher = {Association for Computational Linguistics},
  address   = {Online},
  year      = {2020},
  url       = {https://aclanthology.org/2020.emnlp-main.447}
}

@article{oche2025systematic,
  author    = {Oche, A. J. and Folashade, A. G. and Ghosal, T. and others},
  title     = {A Systematic Review of Key Retrieval-Augmented Generation (RAG) Systems: Progress, Gaps, and Future Directions},
  journal   = {arXiv preprint arXiv:2507.18910},
  year      = {2025},
  url       = {https://arxiv.org/abs/2507.18910}
}

@inproceedings{carlini2021extracting,
  author    = {Nicholas Carlini and Florian Tramer and Eric Wallace and Matthew Jagielski and Ariel Herbert-Voss and Katherine Lee and Adam Roberts and Tom Brown and Dawn Song and Ulfar Erlingsson and others},
  title     = {Extracting Training Data from Large Language Models},
  booktitle = {Proceedings of the 30th USENIX Security Symposium (USENIX Security 21)},
  pages     = {2633--2650},
  year      = {2021},
  publisher = {USENIX Association},
  url       = {https://www.usenix.org/conference/usenixsecurity21/presentation/carlini}
}

@inproceedings{nguyen2016msmarco,
  author    = {Tri Nguyen and Mir Rosenberg and Xia Song and Jianfeng Gao and Saurabh Tiwary and Rangan Majumder and Li Deng},
  title     = {MS MARCO: A Human Generated Machine Reading Comprehension Dataset},
  booktitle = {CoCo@NeurIPS 2016},
  year      = {2016},
  url       = {https://arxiv.org/abs/1611.09268}
}

@inproceedings{chen2023imperceptible,
  author    = {Xuanang Chen and Ben He and Zheng Ye and Le Sun and Yingfei Sun},
  title     = {Towards Imperceptible Document Manipulations against Neural Ranking Models},
  booktitle = {Proceedings of the 61st Annual Meeting of the Association for Computational Linguistics (ACL 2023)},
  year      = {2023},
  publisher = {Association for Computational Linguistics},
  url       = {https://aclanthology.org/2023.acl-long.XXX}
}

@inproceedings{yih2020rag,
  author    = {Scott Wen-tau Yih},
  title     = {Retrieval-Augmented Generation for Knowledge-Intensive NLP Tasks},
  booktitle = {Proceedings of the 34th Conference on Neural Information Processing Systems (NeurIPS 2020)},
  year      = {2020},
  address   = {Vancouver, Canada},
  publisher = {Curran Associates, Inc.}
}

@inproceedings{zou2025poisonedrag,
  author    = {Wei Zou and Xinyu Zhang and Yihan Cao and Yu Zhao and Jiazhao Zhang and Zhiyuan Liu and Maosong Sun and Yang Liu},
  title     = {PoisonedRAG: Knowledge Corruption Attacks to Retrieval-Augmented Generation of Large Language Models},
  booktitle = {Proceedings of the 34th USENIX Security Symposium (USENIX Security 2025)},
  year      = {2025},
  publisher = {USENIX Association}
}

@inproceedings{Liu2024MultiGranular,
  author    = {Yu-An Liu and Rui Zhang and Jiafeng Guo and Maarten de Rijke},
  title     = {Multi-Granular Adversarial Attacks against Black-Box Neural Ranking Models},
  booktitle = {Proceedings of the 47th International ACM SIGIR Conference on Research and Development in Information Retrieval},
  pages     = {1391--1400},
  year      = {2024},
  publisher = {ACM}
}

@inproceedings{Ma2021PROP,
  author    = {Xinyu Ma and Jiafeng Guo and Rui Zhang and Yixing Fan and Xiang Ji and Xueqi Cheng},
  title     = {PROP: Pre-Training with Representative Words Prediction for Ad-Hoc Retrieval},
  booktitle = {Proceedings of the 14th ACM International Conference on Web Search and Data Mining},
  pages     = {283--291},
  year      = {2021},
  publisher = {ACM}
}

@inproceedings{Devlin2019BERT,
  author    = {Jacob Devlin and Ming-Wei Chang and Kenton Lee and Kristina Toutanova},
  title     = {BERT: Pre-training of Deep Bidirectional Transformers for Language Understanding},
  booktitle = {Proceedings of the 2019 Conference of the North American Chapter of the Association for Computational Linguistics: Human Language Technologies},
  year      = {2019}
}

@inproceedings{Chen2022AdversarialNLP,
  author    = {Yangyi Chen and Hongcheng Gao and Ganqu Cui and Fanchao Qi and Longtao Huang and Zhiyuan Liu and Maosong Sun},
  title     = {Why Should Adversarial Perturbations be Imperceptible? Rethink the Research Paradigm in Adversarial NLP},
  booktitle = {Proceedings of the 2022 Conference on Empirical Methods in Natural Language Processing},
  year      = {2022}
}

@inproceedings{Li2025CorpusPoisoning,
  author    = {Yue Li and Panagiotis Eustratiadis and Simon Lupart and Nicola Tonellotto and Maarten de Rijke},
  title     = {Unsupervised Corpus Poisoning Attacks in Continuous Space for Dense Retrieval},
  booktitle = {Proceedings of the 48th International ACM SIGIR Conference on Research and Development in Information Retrieval},
  pages     = {2452--2462},
  year      = {2025},
  publisher = {ACM}
}

@article{Jafarzadeh2015SEAReview,
  author  = {Hamed Jafarzadeh and Ayb{\"u}ke Aurum and John D'Ambra and Mehdi Ghazinoory},
  title   = {A Systematic Review on Search Engine Advertising},
  journal = {Pacific Asia Journal of the Association for Information Systems},
  volume  = {7},
  number  = {3},
  pages   = {2},
  year    = {2015}
}

@article{Chen2020FastDPC,
  author  = {Yong Chen and Xiaohui Hu and Wenjie Fan and Ke Deng},
  title   = {Fast Density Peak Clustering for Large-Scale Data Based on kNN},
  journal = {Knowledge-Based Systems},
  volume  = {187},
  pages   = {104824},
  year    = {2020},
  publisher = {Elsevier}
}

@inproceedings{finocchiaro-et-al:embedding,
  author    = {Jessica Finocchiaro and Rafael Frongillo and Bo Waggoner},
  title     = {An Embedding Framework for Consistent Polyhedral Surrogates},
  booktitle = {Advances in Neural Information Processing Systems},
  year      = {2019}
}

@article{mitchell-lapata-2010,
  author  = {Jeff Mitchell and Mirella Lapata},
  title   = {Composition in Distributional Models of Semantics},
  journal = {Cognitive Science},
  volume  = {34},
  number  = {8},
  pages   = {1388--1429},
  year    = {2010}
}

@inproceedings{ramos-2003-tfidf,
  author    = {Juan Ramos},
  title     = {Using TF-IDF to Determine Word Relevance in Document Queries},
  booktitle = {Proceedings of the First Instructional Conference on Machine Learning},
  year      = {2003}
}

@inproceedings{carlini-2024-poisoning,
  author    = {Nicholas Carlini and Matthew Jagielski and Christopher A. Choquette-Choo and Florian Tramer and Milad Nasr and Katherine Lee and Andreas Terzis and Borja Balle and Nicolas Papernot},
  title     = {Poisoning Web-Scale Training Datasets is Practical},
  booktitle = {Proceedings of the 2024 IEEE Symposium on Security and Privacy (SP)},
  pages     = {407--425},
  year      = {2024},
  publisher = {IEEE}
}

@inproceedings{madry-2018-adversarial,
  author    = {Aleksander Madry and Aleksandar Makelov and Ludwig Schmidt and Dimitris Tsipras and Adrian Vladu},
  title     = {Towards Deep Learning Models Resistant to Adversarial Attacks},
  booktitle = {International Conference on Learning Representations},
  year      = {2018}
}

@inproceedings{holtzman2019curious,
  author    = {Ari Holtzman and Jan Buys and Li Du and Maxwell Forbes and Yejin Choi},
  title     = {The Curious Case of Neural Text Degeneration},
  booktitle = {International Conference on Learning Representations},
  year      = {2020}
}

\clearpage
\appendix
\section{GPT-based Fluency Evaluation Prompt}
\vspace{-10pt}
\begin{figure}[ht]
  \centering
  \includegraphics[width=0.5\textwidth]{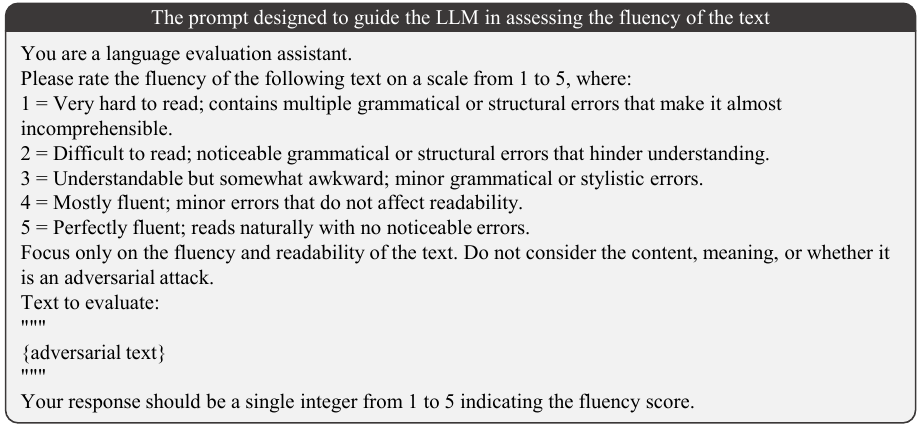}
  \label{app:fig:fluency_template}
\end{figure}
\vspace{-20pt}
\section{NSP Score Analysis of Adversarial vs. Natural Texts}

\begin{figure}[t]
  \centering
  \includegraphics[width=1.0\columnwidth]{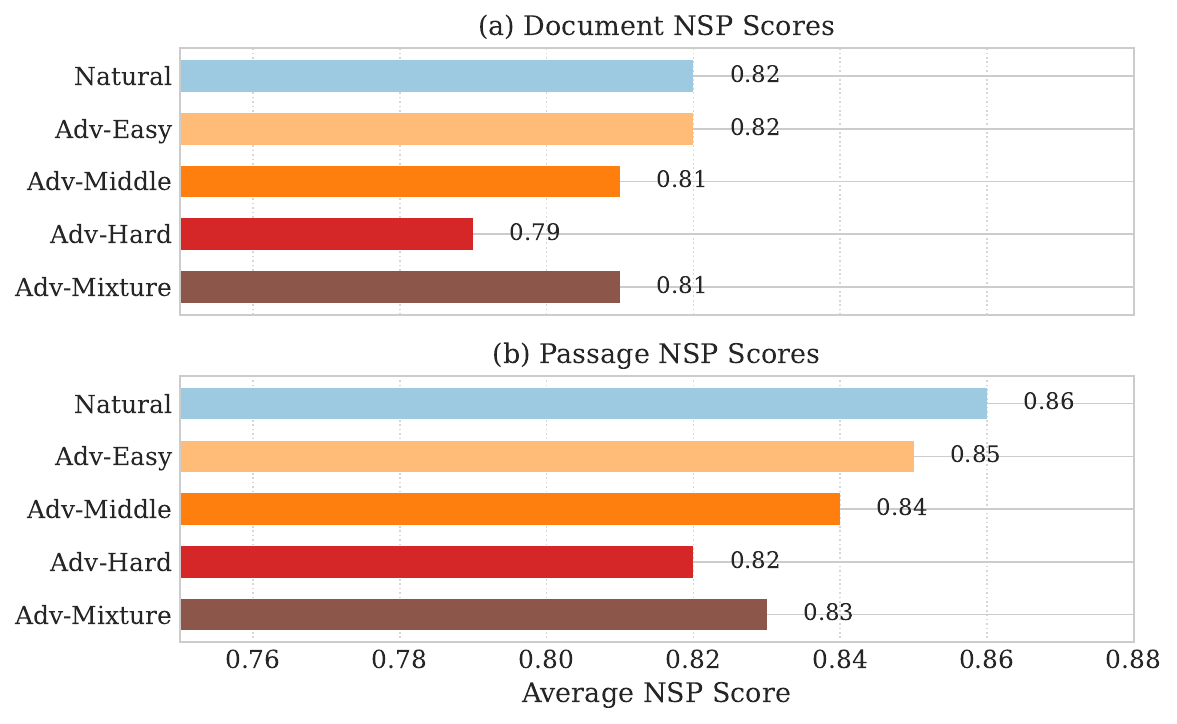}
  \caption{{NSP scores of natural and adversarial texts across difficulty levels.}
  The figure reports the average NSP scores on the MS-MARCO benchmarks, where the top panel corresponds to the Document dataset and the bottom panel corresponds to the Passage dataset.}
  \label{app:fig:nsp_scores}
\end{figure}

Centroid documents that are semantically distant from the target document are more likely to be detected.
To enhance the stealthiness of the concatenation operation, we incorporate a center-guided score into the optimization objective and assess semantic coherence using a Next Sentence Prediction (NSP) model.
We report NSP scores of adversarial and natural texts across four difficulty levels (Easy, Middle, Hard, and Mixture).
Natural sentences are randomly sampled from the corpus as reference baselines, while adversarial texts are generated by our method.
The average NSP scores are computed and visualized in Figure~\ref{app:fig:nsp_scores}.

As shown in Figure~\ref{app:fig:nsp_scores}, although the NSP scores of adversarial texts decrease as the difficulty increases, they remain highly comparable to those of natural texts.
For instance, on the MS-MARCO Document dataset, natural documents achieve an NSP score of 0.82, while adversarial documents at the Mixture level obtain a score of 0.81, exhibiting only a marginal difference.
These results indicate that the NSP model continues to regard the concatenated centroid documents as valid subsequent sentences, suggesting that the concatenation operation is not detected by NSP-based coherence checks.

\section{Evaluation of Surrogate Model Training Strategies}
We evaluate the effectiveness of different surrogate model training strategies on Mixture-level targets (Table~\ref{app:tab:wh_com}). 
During training, the Iterative Retrieval Module (IRM) constructs the training dataset by iteratively retrieving relevant documents.  
SentAttack$_{random}$ does not use IRM and relies solely on documents from other queries as negatives.
As shown in Table~\ref{app:tab:wh_com}, SentAttack consistently outperforms SentAttack$_{random}$ (SRR@10 35.2 vs.\ 32.1; SRR@100 32.7 vs.\ 29.4) and approaches the performance of SentAttack$_{white}$.  
SentAttack$_{random}$, which does not leverage IRM, fails to gather sufficient retrieval information from the limited outputs of the black-box RAG system, resulting in lower performance compared to SentAttack.

\begin{table}[!t]
  \centering
  \setlength{\tabcolsep}{4pt}
  \resizebox{\columnwidth}{!}{
  \begin{tabular}{|l|S[table-format=2.1]S[table-format=2.1]|S[table-format=2.1]
                  |S[table-format=2.1]S[table-format=2.1]|S[table-format=2.1]|}
    \toprule
    & \multicolumn{3}{c|}{MS-MARCO Document} 
    & \multicolumn{3}{c|}{MS-MARCO Passage} \\
    \cmidrule(lr){2-4} \cmidrule(lr){5-7}
    \multicolumn{1}{|c|}{Method}
    & \multicolumn{2}{c|}{SRR} & NRS
    & \multicolumn{2}{c|}{SRR} & NRS \\
    \cmidrule(lr){2-4} \cmidrule(lr){5-7}
           & {@10} & {@100} & {@100}
           & {@100} & {@1000} & {@1000} \\
    \midrule
    SentAttack               & 35.2 & 74.3 & 71.5 & 32.7 & 65.3 & 63.2 \\
    SentAttack$_{random}$    & 32.1 & 72.5 & 68.8 & 29.4 & 63.7 & 62.1 \\
    SentAttack$_{white}$     & 38.8 & 77.1 & 74.2 & 34.5 & 68.1 & 65.4 \\
    \bottomrule
  \end{tabular}}
  \caption{Effectiveness of different surrogate model training strategies on Mixture-level targets in MS-MARCO Document and Passage Ranking Dataset.}
  \label{app:tab:wh_com}
\end{table}

\begin{table*}[!t]
	\centering
	\resizebox{\textwidth}{!}{
	\begin{tabular}{l|l|ccc|ccc|ccc|ccc}
	  \toprule
	  &
	  & \multicolumn{3}{c|}{Easy} 
	  & \multicolumn{3}{c|}{Middle} 
	  & \multicolumn{3}{c|}{Hard} 
	  & \multicolumn{3}{c}{Mixture} \\
	  \cmidrule(lr){3-5} \cmidrule(lr){6-8} \cmidrule(lr){9-11} \cmidrule(lr){12-14}
	  Dataset & \multicolumn{1}{l|}{Method} 
	  & \multicolumn{2}{c}{SRR} & NRS
	  & \multicolumn{2}{c}{SRR} & NRS
	  & \multicolumn{2}{c}{SRR} & NRS
	  & \multicolumn{2}{c}{SRR} & NRS \\
	  \cmidrule(lr){3-14}
	  & & @10 & @100 & @100
		& @10 & @100 & @100
		& @10 & @100 & @100
		& @10 & @100 & @100 \\
	  \midrule
	  \multirow{7}{*}{\makecell[l]{MS-MARCO\\Document}}
	  & TF-IDF  & 16.0 & 40.9 & 32.1 & 11.1 & 28.0 & 23.6 & 4.2 & 14.4 & 13.6 & 10.3 & 28.6 & 23.2 \\
	  & TS      & 37.8 & 88.1 & 67.5 & 27.2 & 58.0 & 60.3 & 15.1 & 35.8 & 33.5 & 23.0 & 58.0 & 53.8 \\
	  & PAT     & 26.5 & 70.2 & 52.2 & 13.7 & 36.0 & 32.0 & 7.9  & 27.1 & 26.4 & 12.7 & 44.5 & 37.8 \\
	  & PRADA   & 28.4 & 74.7 & 56.2 & 18.5 & 43.1 & 37.9 & 11.2 & 33.0 & 33.3 & 19.3 & 50.3 & 44.1 \\
	  & MCARA   & 43.5 & 92.3 & 73.1 & 28.1 & 66.5 & 61.4 & 24.4 & 50.2 & 51.3 & 32.0 & 72.0 & 62.0 \\
	  \cmidrule(lr){2-14}
	  & SentAttack$_{noit}$
				& \textbf{55.8} & 93.2 & 85.4
				& 50.7 & 83.5 & 81.8
				& 33.5 & 56.2 & 54.6
				& 36.0 & 74.8 & 72.1 \\
	  & SentAttack
				& 55.4 & \textbf{93.7} & \textbf{85.6}
				& \textbf{51.9} & \textbf{84.1} & \textbf{82.3}
				& \textbf{34.9} & \textbf{57.5} & \textbf{55.7}
				& \textbf{36.1} & \textbf{75.4} & \textbf{72.8} \\
	  \midrule
	  & & @100 & @1000 & @1000
		& @100 & @1000 & @1000
		& @100 & @1000 & @1000
		& @100 & @1000 & @1000 \\
	  \midrule
	  \multirow{7}{*}{\makecell[l]{MS-MARCO\\Passage}}
	  & TF-IDF  & 10.2 & 35.2 & 25.1 & 6.4  & 19.8 & 18.3 & 2.1  & 10.5 & 10.3 & 6.1  & 21.6 & 17.8 \\
	  & TS      & 28.6 & 79.0 & 59.1 & 17.2 & 50.8 & 48.7 & 8.4  & 27.6 & 26.9 & 17.8 & 52.0 & 44.4 \\
	  & PAT     & 16.4 & 62.3 & 46.7 & 9.4  & 30.0 & 28.6 & 5.3  & 23.4 & 21.5 & 10.4 & 38.6 & 32.3 \\
	  & PRADA   & 28.4 & 68.2 & 51.0 & 13.8 & 39.9 & 39.6 & 10.6 & 31.5 & 30.1 & 14.7 & 46.4 & 40.2 \\
	  & MCARA   & 20.1 & 83.1 & 65.9 & 22.7 & 57.3 & 53.7 & 15.3 & 41.1 & 40.2 & 23.7 & 60.5 & 53.3 \\
	  \cmidrule(lr){2-14}
	  & SentAttack$_{noit}$
				& \textbf{51.0} & 92.3 & 83.0
				& \textbf{46.2} & 82.5 & 80.0
				& 30.8 & 54.2 & 52.7
				& \textbf{34.5} & {64.8} & {62.4} \\
	  & SentAttack
				& {50.6} & \textbf{92.8} & \textbf{83.2}
				& {45.9} & \textbf{83.2} & \textbf{80.5}
				& \textbf{31.2} & \textbf{54.9} & \textbf{53.3}
				& {33.9} & \textbf{65.3} & \textbf{63.1} \\
	  \bottomrule
	\end{tabular}}
	\caption{Performance comparison of different adversarial attack methods on MS-MARCO Document and Passage datasets. Best results in each column are highlighted in \textbf{bold}. SentAttack$_{\mathit{noit}}$ denotes the variant without iterative retrieval.}
	\label{app:tab:attack_results}
  \end{table*}

\section{Impact of Retrieval Depth on Attack Performance}
\subsection{Retrieval Depth in RAG Systems}

RAG systems rely on a dense retriever to return a compact set of highly relevant documents per query. 
The number of retrieved documents, $K$, directly affects retrieval accuracy and downstream generation quality.
In practice, $K$ is often small ($K \in [5,10]$) due to latency, memory, and LLM input-length constraints~\cite{yih2020rag}.
Consequently, only a concise evidence set reaches the generator, limiting the information available for adversarial manipulation.

These retrieval depths align with common experimental settings in both RAG and recent RAG-attack studies.
For instance, retrieves top-$K=5$ documents using Contriever and GPT-4o-mini, while \cite{zou2025poisonedrag} retrieves top-$5$ documents from NQ with Contriever.
Classical RAG studies~\cite{yih2020rag} also adopt $K \in \{5,10\}$, reflecting practical deployment constraints and serving as a standard setting for evaluating dense retriever robustness.

\subsection{Retrieval Depths in Existing Attack Baselines}

Many adversarial attacks on dense retrievers or neural ranking models adopt retrieval settings that deviate from practical RAG deployments.
While real-world RAG systems typically retrieve only a small number of documents ($K \in [5,10]$), some state-of-the-art attacks use substantially larger retrieval depths, such as $K=100$ for MS-MARCO Document and $K=1000$ for MS-MARCO Passage~\cite{liu2023blackbox,wu2023prada}.
These settings implicitly assume access to excessively large candidate pools, which may overestimate attack effectiveness under realistic RAG deployment constraints.
While such configurations are useful for analyzing upper-bound attack capabilities, they do not reflect the retrieval limitations faced by deployed RAG systems.
To better reflect practical conditions, we adopt a constrained retrieval depth of $K=10$, enabling a more realistic and reliable assessment of adversarial attack effectiveness.

\begin{figure}[t]
  \centering
  \begin{subfigure}[t]{0.48\columnwidth}
    \centering
    \includegraphics[width=\linewidth,trim=5 5 5 5,clip]{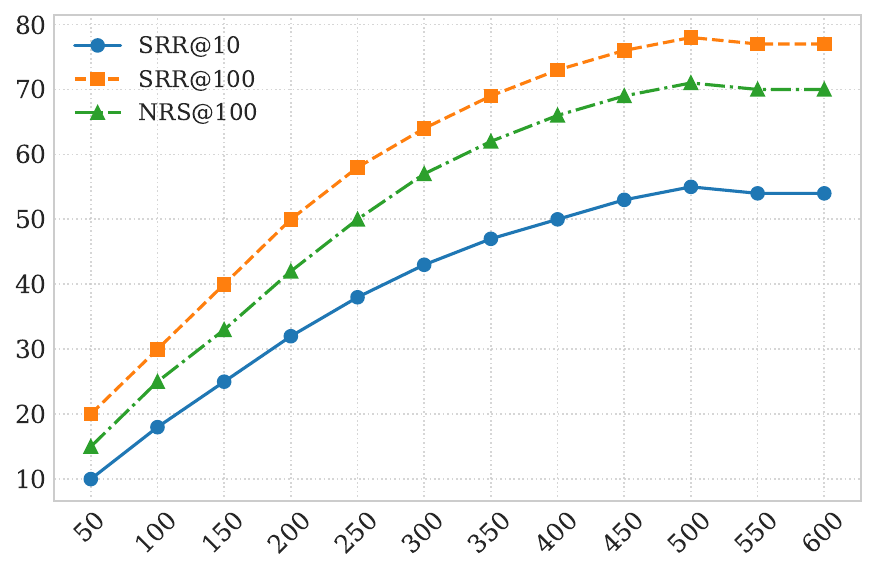}
    \caption{MARCO Document}
    \label{app:fig:attack_document}
  \end{subfigure}
  \hfill
  \begin{subfigure}[t]{0.48\columnwidth}
    \centering
    \includegraphics[width=\linewidth,trim=5 5 5 5,clip]{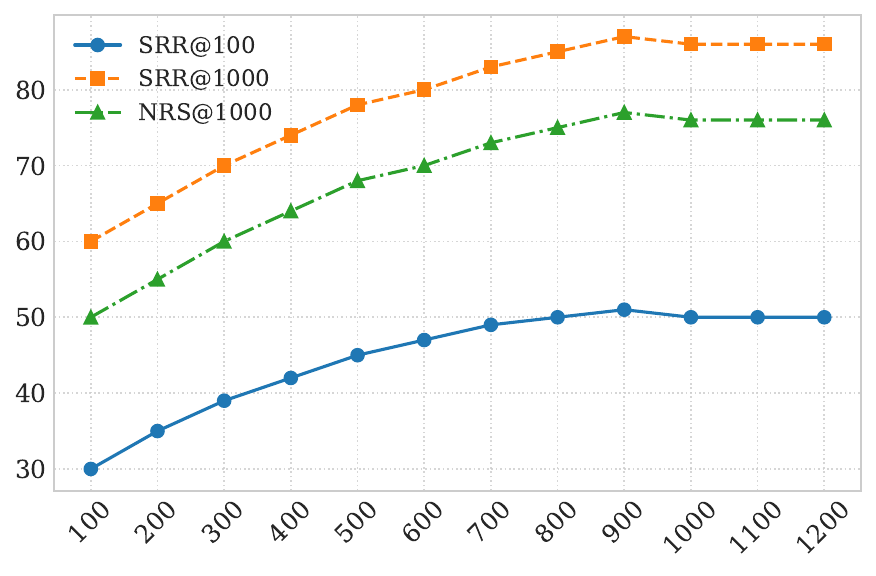}
    \caption{MARCO Passage}
    \label{app:fig:attack_passage}
  \end{subfigure}

  \caption{{SentAttack performance with varying numbers of retrieved relevant texts.} Each curve shows one evaluation metric.}
  \label{app:fig:attack_performance_retrieved_texts}
\end{figure}

\subsection{Effectiveness Under Small-$K$ Retrieval}

Most existing adversarial attacks on dense retrievers assume access to a large and fixed top-$K$ candidate pool (e.g., $K=100$ for MS-MARCO Document and $K=1000$ for MS-MARCO Passage).
To reproduce these settings, our evaluation includes a variant, SentAttack$_{\mathit{noit}}$, which directly performs the attack under these large-$K$ configurations without iterative retrieval module (IRM).
That is, SentAttack$_{\mathit{noit}}$ denotes a variant of SentAttack without the iterative retrieval mechanism.

Table~\ref{app:tab:attack_results} presents a systematic comparison of different attack methods on the MS-MARCO Document and Passage datasets.
The results show that SentAttack$_{\mathit{noit}}$ and the full SentAttack model achieve comparable overall performance.
This indicates that the iterative retrieval mechanism in SentAttack can effectively simulate large candidate pool configurations, allowing the attack to accumulate large-scale retrieval information starting from a realistic small-$K$ setting.

\subsection{Effect of Iteratively Retrieved Document Count on Attack Performance}

Figure~\ref{app:fig:attack_performance_retrieved_texts} examines the impact of the number of iteratively retrieved relevant texts on attack performance.
Specifically, we evaluate the effectiveness of Mixture-level adversarial texts under varying amounts of retrieved texts on both MS-MARCO Document and Passage datasets.

As shown in the figure, the attack performance generally improves as more relevant texts are obtained through iterative retrieval.
When the number of retrieved documents approaches dataset-specific saturation points (around 500 for MS-MARCO Document and 900--1000 for MS-MARCO Passage), performance reaches its peak and then stabilizes or slightly declines, indicating diminishing returns from introducing additional documents.
These results demonstrate that the iterative retrieval mechanism can effectively approximate direct access to large top-$K$ candidate sets, and is a key factor enabling SentAttack to achieve strong performance under realistic RAG constraints.

\section{Theoretical proof}
\label{app:appendix:center_theory}

We provide a theoretical proof for why SenAttack remains effective
under a surrogate-to-target setting, where the attacker optimizes adversarial documents against a
surrogate retrirver that only approximates the target retriever.
Let $d_t$ denote the target document and $d_{c_i}$ a centroid document,
whose surrogate embedding is $c_i = E_s(d_{c_i})$.
Here, $E_s(\cdot)$ denotes the document encoder of the surrogate model,
while $E_t(\cdot)$ denotes the document encoder of the target model.

\subsection{Assumptions}

\paragraph{Assumption 1: Embedding Discrepancy Bound.}  
For any query or document \(x\), the embeddings of the target model and surrogate model satisfy
\begin{equation}
\| E_t(x) - E_s(x) \|_2 \le \delta,
\end{equation}
ensuring that the surrogate embedding is a \(\delta\)-close approximation of the target embedding (as assumed in \cite{finocchiaro-et-al:embedding}).  

\paragraph{Proposition 1: Surrogate–Target Similarity Score Transfer.}  
Under assumption 1, the similarity score between a query \(q\) and a document \(d\) under the target model
is close to that under the surrogate model:
\begin{equation}
\begin{split}
&\big| \langle E_t(q), E_t(d) \rangle - \langle E_s(q), E_s(d) \rangle \big| \\
&\qquad \le \delta \|E_t(d)\|_2 + \delta \|E_s(q)\|_2.
\end{split}
\end{equation}

\paragraph{Proof.}
\begin{equation}
\begin{split}
&\big| \langle E_t(q), E_t(d) \rangle - \langle E_s(q), E_s(d) \rangle \big| \\
&= \big| \langle E_t(q)-E_s(q), E_t(d) \rangle \\
&\quad + \langle E_s(q), E_t(d)-E_s(d) \rangle \big| \\
&\le \| E_t(q)-E_s(q) \|_2 \, \| E_t(d) \|_2 \\
&\quad + \| E_s(q) \|_2 \, \| E_t(d)-E_s(d) \|_2 \\
&\le \delta \|E_t(d)\|_2 + \delta \|E_s(q)\|_2
\end{split}
\end{equation}

If the query and document embeddings are $L_2$-normalized (\(\|E_s(q)\|_2 = \|E_t(d)\|_2 = 1\)), the bound simplifies to
\begin{equation}
\big| \langle E_t(q), E_t(d) \rangle - \langle E_s(q), E_s(d) \rangle \big| \le 2 \delta.
\end{equation}

\paragraph{Assumption 2: Compositionality of the Surrogate Encoder.}
For text concatenation $u \oplus v$, the surrogate encoder satisfies
\begin{equation}
\begin{split}
E_s(u \oplus v) &= \alpha_u E_s(u) + \alpha_v E_s(v) + \epsilon_{\mathrm{comp}}, \\
& \alpha_u,\alpha_v \ge 0,\ \alpha_u+\alpha_v=1.
\end{split}
\end{equation}
This reflects the empirical observation that the embedding of the concatenated text $u \oplus v$ can be approximated by a convex combination of the embeddings of $u$ and $v$,
where $\alpha_u$ and $\alpha_v$ correspond to the proportion of words in $u$ and $v$ relative to the total length. 
We assume that the approximation error is bounded, i.e., $\|\epsilon_{\mathrm{comp}}\|_2 \le \eta_{\mathrm{comp}}$. 
Following \cite{mitchell-lapata-2010}, we use the embeddings produced by the retrieval model to represent the semantic content of the text segments,
so that the semantic embedding of the concatenated text can be expressed as a combination of the embeddings of its components, with $\epsilon_{\mathrm{comp}}$ and $\eta_{\mathrm{comp}}$ quantifying the approximation error.

\subsection{Concatenated Initialization}
Define the concatenated initial document
\begin{equation}
\tilde{d}_i^{(0)} := d_t \oplus d_{c_i},
\end{equation}
with embeddings
\begin{equation}
\boldsymbol{e}_{\tilde{d}_i^{(0)}}^{(s)} := E_s(\tilde{d}_i^{(0)}), \qquad
\boldsymbol{e}_{d_t}^{(s)} := E_s(d_t).
\end{equation}
Here $\tilde d_i^{(0)}$ denotes the initialization formed by appending
the centroid document $d_{c_i}$ to the target document $d_t$.

\paragraph{Deriving the Concatenated Representation.}
Applying Assumption 2 for $u=d_t$ and $v=d_{c_i}$, we obtain
\begin{equation}
\boldsymbol{e}_{\tilde{d}_i^{(0)}}^{(s)}
= (1-\alpha_i)\boldsymbol{e}_{d_t}^{(s)} + \alpha_i c_i + \epsilon_{\mathrm{comp}},
\end{equation}
where $c_i := E_s(d_{c_i})$ and $\alpha_i := \alpha_{d_{c_i}}$.  
Intuitively, $\alpha_i$ measures the contribution of the centroid document
to the overall representation in the surrogate embedding space.
This expresses the concatenated embedding as an approximate convex combination
of the target-document embedding and the centroid-document embedding,
with the approximation error explicitly given by $\epsilon_{\mathrm{comp}}$, which is bounded by $\eta_{\mathrm{comp}}$.

\subsection{Initial Gain from Concatenation}

Define the similarity score gain under the surrogate encoder:
\begin{equation}
G_i^{(0)} = 
\langle \boldsymbol{e}_q^{(s)}, 
\boldsymbol{e}_{\tilde{d}_i^{(0)}}^{(s)} - \boldsymbol{e}_{d_t}^{(s)}
\rangle.
\end{equation}
This quantity measures how much the surrogate-model similarity score for $d_t$ increases immediately after concatenating centroid document $d_{c_i}$.

\paragraph{Derivation.}
Substituting the compositional form:
\begin{equation}
\begin{split}
G_i^{(0)} 
&= \alpha_i \langle \boldsymbol{e}_q^{(s)}, c_i - \boldsymbol{e}_{d_t}^{(s)} \rangle
 + \langle \boldsymbol{e}_q^{(s)}, \epsilon_{\mathrm{comp}} \rangle.
\end{split}
\end{equation}
The first term captures alignment-driven gain, proportional to $\alpha_i$, while the second term is the composition error contribution.  
Bounding the error term using Cauchy--Schwarz yields
\begin{equation}
\big|\langle \boldsymbol{e}_q^{(s)}, \epsilon_{\mathrm{comp}} \rangle\big|
\le \|\boldsymbol{e}_q^{(s)}\|_2 \eta_{\mathrm{comp}},
\end{equation}
which implies the lower bound
\begin{equation}
G_i^{(0)} 
\ge \alpha_i \langle \boldsymbol{e}_q^{(s)}, c_i - \boldsymbol{e}_{d_t}^{(s)} \rangle
 - \|\boldsymbol{e}_q^{(s)}\|_2 \eta_{\mathrm{comp}}.
\end{equation}
Thus, under the standard assumption that all query and document embeddings
are $L_2$-normalized, i.e.,
$\|\boldsymbol{e}_q^{(s)}\|_2 = \|c_i\|_2 = \|\boldsymbol{e}_{d_t}^{(s)}\|_2 = 1$,
the lower bound simplifies to
\begin{equation}
G_i^{(0)}
\ge
\alpha_i \left\langle \boldsymbol{e}_q^{(s)},
c_i - \boldsymbol{e}_{d_t}^{(s)} \right\rangle
- \eta_{\mathrm{comp}} .
\end{equation}
Each centroid document $c_i$ is more aligned with the target
query than the original document $d_t$ in the surrogate embedding space, which
implies
\(
\langle \boldsymbol{e}_q^{(s)}, c_i \rangle
\ge
\langle \boldsymbol{e}_q^{(s)}, \boldsymbol{e}_{d_t}^{(s)} \rangle
\).
Therefore, the first term is non-negative.
As a result, in the worst case, the initial surrogate gain introduced by concatenation is lower bounded by the composition
error $\eta_{\mathrm{comp}}$, indicating that the concatenation operation does
not introduce detrimental effects beyond a bounded composition noise.

\subsection{Effect of Adversarial Perturbation (synonym substituting)}

After concatenation we add a discrete perturbation $p$ to further optimize the adversarial document:
\begin{equation}
\tilde{d}_i^{b} = \tilde{d}_i^{(0)} \oplus p,
\end{equation}
where $\tilde{d}_i^{b}$ denotes the adversarial document after discrete perturbation, 
and $p$ is constrained by a budget (e.g., edit count or norm).

\paragraph{Query-guided surrogate objective.}
Recall that in our method the query-guided component of the surrogate objective is
\begin{equation}
\mathcal{O}_{\mathrm{query}}(\tilde{d}_i, q) =
\frac{
\boldsymbol{e}_q^{(s)\top} \, E_s(\tilde{d}_i)
}{
\|\boldsymbol{e}_q^{(s)}\|_2 \, \|E_s(\tilde{d}_i)\|_2
}.
\end{equation}
The perturbation $p$ is selected to maximize this objective within the allowed budget:
\begin{equation}
\tilde{d}_i^{b} = \arg\max_{\tilde{d} \in \mathcal{B}(\tilde{d}_i^{(0)})}
\mathcal{O}_{\mathrm{query}}(\tilde{d}, q),
\end{equation}
where $\mathcal{B}(\tilde{d}_i^{(0)})$ denotes the set of documents reachable
from $\tilde{d}_i^{(0)}$ under the allowed perturbation.

\paragraph{Derivation of a Lower Bound on the Perturbation Gain.}
Let the surrogate-model similarity score gain due to $p$ be
\begin{equation}
\Delta s_s(p) :=
\langle \boldsymbol{e}_q^{(s)}, E_s(\tilde{d}_i^{b}) - E_s(\tilde{d}_i^{(0)}) \rangle.
\end{equation}
By construction, each token-level edit in $p$ is accepted only if it does not decrease
the query-guided objective. Therefore, we have
\begin{equation}
\Delta s_s(p) \ge 0,
\end{equation}
providing a positive lower bound for the gain in the surrogate embedding space.

\paragraph{Total Surrogate Gain.}
The total gain obtained by the adversarial document $\tilde{d}_i^{b}$
in the surrogate model is
\begin{equation}
\Delta s_s^{\mathrm{total}} = G_i^{(0)} + \Delta s_s(p),
\end{equation}
where $G_i^{(0)}$ is the initial gain from concatenating the centroid document to $d_t$, 
and $\Delta s_s(p)$ is the additional gain obtained by discrete perturbation. 
Thus, the total surrogate-model gain satisfies
\begin{equation}
\Delta s_s^{\mathrm{total}} \ge G_i^{(0)},
\end{equation}
showing that the total surrogate-model similarity gain
resulting from the concatenation-plus-perturbation procedure
is always greater than or equal to the gain induced by concatenation alone.

\subsection{Lower Bound on Target-Model Improvement}

We now transfer the gain guarantee from the surrogate model to the target model.
By proposition~1, the similarity score under the target model is lower bounded by
its surrogate counterpart up to a bounded discrepancy:
\begin{equation}
\begin{split}
\Delta s_t
&= \big\langle E_t(q), E_t(\tilde{d}_i^b) - E_t(d_t) \big\rangle \\
&\ge \big\langle E_s(q), E_s(\tilde{d}_i^b) - E_s(d_t) \big\rangle - 2\delta \\
&= \Delta s_s^{\mathrm{total}} - 2\delta ,
\end{split}
\end{equation}
where $\tilde{d}_i^b$ denotes the optimized adversarial document and
$\delta$ is the surrogate-to-target discrepancy bound.

Recall that the total surrogate-model gain decomposes as
\begin{equation}
\Delta s_s^{\mathrm{total}} = G_i^{(0)} + \Delta s_s(p),
\end{equation}
where $\Delta s_s(p) \ge 0$ due to the query-guided optimization.
From the analysis of the concatenated initialization, we have
\begin{equation}
G_i^{(0)}
\ge
\alpha_i \big\langle \boldsymbol{e}_q^{(s)}, c_i - \boldsymbol{e}_{d_t}^{(s)} \big\rangle
- \eta_{\mathrm{comp}} .
\end{equation}

Combining the above results yields the final lower bound:
\begin{equation}
\begin{split}
\Delta s_t
\ge\;
&\alpha_i \big\langle \boldsymbol{e}_q^{(s)}, c_i - \boldsymbol{e}_{d_t}^{(s)} \big\rangle \\
&- \eta_{\mathrm{comp}} - 2\delta .
\end{split}
\end{equation}

\paragraph{Analysis of the lower bound.}
The derived lower bound shows that the dominant contribution to the target-model gain
comes from the term
$\alpha_i \langle \boldsymbol{e}_q^{(s)}, c_i - \boldsymbol{e}_{d_t}^{(s)} \rangle$.
This term quantifies the semantic advantage of the centroid document $c_i$
over the original document $d_t$ with respect to the query in the surrogate embedding space,
scaled by the effective contribution $\alpha_i$ of the concatenated centroid.

When the selected centroid is more query-aligned than the original document,
$\langle \boldsymbol{e}_q^{(s)}, c_i - \boldsymbol{e}_{d_t}^{(s)} \rangle$ is positive,
and a non-negligible $\alpha_i$ ensures that this advantage is preserved after concatenation.
In this case, the lower bound remains positive up to the composition error
$\eta_{\mathrm{comp}}$ and the surrogate--target discrepancy $2\delta$.

\section{Case Study}

To demonstrate how SentAttack constructs adversarial documents in practice, we present three case studies from the MS-MARCO Passage dev set for the query \textit{``hillsborough community college president''}, covering easy, middle, and hard target documents. 
The difficulty levels are defined based on the initial ranking of the target document: easy cases correspond to documents initially ranked in the range of [1000, 2000], middle cases are ranked in [2000, 10000], and hard cases are ranked outside the top 10000.

For all three target documents, the same query is used. To improve experimental efficiency, the iterative retrieval step is performed only once per query. 
As a result, the three case studies share the same set of retrieved documents and the same centroid documents, even though their target documents differ. 
These centroid documents typically consist of dictionary-style or definitional passages (e.g., explanations of the verb ``linger'').

When concatenated with the centroid documents, these original target documents immediately shift the document representation toward the semantic region of the query, leading to substantial ranking changes. Building on this concatenation, SentAttack further applies minimal word-level synonym substitutions to generate the final adversarial document. 
In Tables~\ref{app:tab:easy_case}, \ref{app:tab:middle_case}, and \ref{app:tab:hard_case}, the concatenated centroid documents are highlighted in red, while synonym substitutions are highlighted in green.

\begin{table*}[!t]
	\centering
	\begin{tabular}{p{0.16\linewidth} p{0.72\linewidth} p{0.08\linewidth}}
	\toprule
	\textbf{Stage} & \textbf{Content} & \textbf{Rank} \\
	\midrule
  
	\textbf{Original Target Document} &
	What does GRIP stand for?
	Note: We have 100 other definitions for GRIP in our Acronym Attic.
	new search; suggest new definition;
	Search for GRIP in Online Dictionary Encyclopedia.
	&
	1025 \\
	\midrule
  
	\multicolumn{3}{l}{\textbf{Stage 1: Concatenation of Target Document with Different Centroid Documents (highlighted in red)}} \\
	\midrule
  
	\textbf{Centroid 1} &
	What does GRIP stand for?
	Note: We have 100 other definitions for GRIP in our Acronym Attic.
	new search; suggest new definition;
	Search for GRIP in Online Dictionary Encyclopedia.
	\textcolor{red}{linger -- definition and synonyms
	This is the British English definition of linger.
	View American English definition of linger.
	Change your default dictionary to American English.
	View the pronunciation for linger.}
	&
	2 \\
	\midrule
  
	\textbf{Centroid 2} &
	What does GRIP stand for?
	Note: We have 100 other definitions for GRIP in our Acronym Attic.
	new search; suggest new definition;
	Search for GRIP in Online Dictionary Encyclopedia.
	\textcolor{red}{linger
	linger.
	1 to remain or stay in a place longer than usual.
	2 to remain alive or persist.}
	&
	6662 \\
	\midrule
  
	\textbf{Centroid 3} &
	What does GRIP stand for?
	Note: We have 100 other definitions for GRIP in our Acronym Attic.
	new search; suggest new definition;
	Search for GRIP in Online Dictionary Encyclopedia.
	\textcolor{red}{imposed
	imposed refers to something unwelcome or unpleasant that must be endured.}
	&
	1339 \\
	\midrule
  
	\multicolumn{3}{l}{\textbf{Stage 2: Optimized Adversarial Document (synonym substitutions highlighted in green)}} \\
	\midrule
  
	\textbf{Final Adversarial Document} &
	what does \textcolor{darkgreen}{grasping} stand for?
	note: we have 100 other definitions for grip in our acronym attic.
	new search; suggest new definition;
	search for grip in online dictionary encyclopedia.
	\textcolor{red}{linger -- definition and synonyms
	This is the British English definition of linger.
	View American English definition of linger.
	Change your default \textcolor{darkgreen}{vocabulary} to American English.
	View the pronunciation for linger.}
	&
	18 \\
	\bottomrule
	\end{tabular}
  
	\caption{\textbf{Easy-level case study for the query ``hillsborough community college president.''} 
	The target document 
	``What does GRIP stand for? Note: We have 100 other definitions for GRIP in our Acronym Attic; new search; suggest new definition; Search for GRIP in Online Dictionary Encyclopedia'' 
	is initially ranked 1025. 
	After concatenation with centroid documents, the initial candidate rankings vary widely (2, 6662, 1339), 
	and the best candidate is further refined through synonym substitutions, 
	resulting in a final ranking of 18.}

	\label{app:tab:easy_case}
\end{table*}
\clearpage

\begin{table*}[!t]
	\centering
	\begin{tabular}{p{0.16\linewidth} >{\raggedright\arraybackslash}p{0.72\linewidth} p{0.08\linewidth}}
	\toprule
	\textbf{Stage} & \textbf{Content} & \textbf{Rank} \\
	\midrule
  
	\textbf{Original Target Document} &
	Information provided about Plaza: Plaza meaning in Hindi: Get detailed meaning of PLAZA in Hindi language. 
	This page shows Plaza meaning in Hindi with Plaza definition, translation and usage. 
	This page provides translation and definition of Plaza in Hindi language along with grammar, synonyms and antonyms.
	&
	8067 \\
	\midrule
  
	\multicolumn{3}{l}{\textbf{Stage 1: Concatenation of Target Document with Different Centroid Documents (highlighted in red)}} \\
	\midrule
  
	\textbf{Centroid 1} &
	Information provided about Plaza: Plaza meaning in Hindi: Get detailed meaning of PLAZA in Hindi language. 
	This page shows Plaza meaning in Hindi with Plaza definition, translation and usage. 
	This page provides translation and definition of Plaza in Hindi language along with grammar, synonyms and antonyms.
	\textcolor{red}{linger -- definition and synonyms
	This is the British English definition of linger.
	View American English definition of linger.
	Change your default dictionary to American English.
	View the pronunciation for linger.}
	&
	243 \\
	\midrule
  
	\textbf{Centroid 2} &
	Information provided about Plaza: Plaza meaning in Hindi: Get detailed meaning of PLAZA in Hindi language. 
	This page shows Plaza meaning in Hindi with Plaza definition, translation and usage. 
	This page provides translation and definition of Plaza in Hindi language along with grammar, synonyms and antonyms.
	\textcolor{red}{linger
	linger.
	1 to remain or stay in a place longer than usual.
	2 to remain alive or persist.}
	&
	\(> 10000\)  \\
	\midrule
  
	\textbf{Centroid 3} &
	Information provided about Plaza: Plaza meaning in Hindi: Get detailed meaning of PLAZA in Hindi language. 
	This page shows Plaza meaning in Hindi with Plaza definition, translation and usage. 
	This page provides translation and definition of Plaza in Hindi language along with grammar, synonyms and antonyms.
	\textcolor{red}{imposed
	imposed refers to something unwelcome or unpleasant that must be endured.}
	&
	4503 \\
	\midrule
  
	\multicolumn{3}{l}{\textbf{Stage 2: Optimized Adversarial Document (synonym substitutions highlighted in green)}} \\
	\midrule
  
	\textbf{Final Adversarial Document} &
	Information provided about Plaza: Plaza meaning in \textcolor{darkgreen}{Minneapolis}: Get detailed meaning of Plaza in \textcolor{darkgreen}{Asian} language. 
	This page shows Plaza meaning in \textcolor{darkgreen}{Maha} with Plaza definition, translation and usage. 
	This page provides translation and definition of Plaza in \textcolor{darkgreen}{Kumar} language along with grammar, synonyms and antonyms.
	\textcolor{red}{linger -- definition and synonyms
	This is the British English definition of linger.
	View American English definition of linger.
	Change your default \textcolor{darkgreen}{vocabulary} to American English.
	View the \textcolor{darkgreen}{spelled} for linger.}
	&
	1 \\
  
	\bottomrule
	\end{tabular}
  
	\caption{\textbf{Middle-level case study for the query ``hillsborough community college president.''} 
	The target document 
	``Information provided about Plaza: Plaza meaning in Hindi: Get detailed meaning of PLAZA in Hindi language. This page shows Plaza meaning in Hindi with Plaza definition, translation and usage. This page provides translation and definition of Plaza in Hindi language along with grammar, synonyms and antonyms'' 
	is initially ranked 8067. 
	After concatenation with centroid documents, the initial candidate rankings vary widely (243, outside top-10000
	, 4503), 
	and the best candidate is further refined through synonym substitutions, 
	resulting in a final ranking of 1.}

	\label{app:tab:middle_case}
\end{table*}

\clearpage
\begin{table*}[!t]
\centering
\begin{tabular}{p{0.16\linewidth} >{\raggedright\arraybackslash}p{0.72\linewidth} p{0.08\linewidth}}
\toprule
\textbf{Stage} & \textbf{Content} & \textbf{Rank} \\
\midrule

\textbf{Original Target Document} &
There is a reason why it makes sense to use 12 am for midnight: because it’s before noon of the same day. 
This reinforces the idea that the stroke of midnight begins the new day. 
As for 12 pm, it’s true that it’s at noon, not after it, but wait!
&
\textgreater 10000 \\
\midrule

\multicolumn{3}{l}{\textbf{Stage 1: Concatenation of Target Document with Different Centroid Documents (highlighted in red)}} \\
\midrule

\textbf{Centroid 1} &
There is a reason why it makes sense to use 12 am for midnight: because it’s before noon of the same day. 
This reinforces the idea that the stroke of midnight begins the new day. 
As for 12 pm, it’s true that it’s at noon, not after it, but wait!
\textcolor{red}{linger -- definition and synonyms
This is the British English definition of linger.
View American English definition of linger.
Change your default dictionary to American English.
View the pronunciation for linger.}
&
243 \\
\midrule

\textbf{Centroid 2} &
There is a reason why it makes sense to use 12 am for midnight: because it’s before noon of the same day. 
This reinforces the idea that the stroke of midnight begins the new day. 
As for 12 pm, it’s true that it’s at noon, not after it, but wait!
\textcolor{red}{linger
linger.
1 to remain or stay in a place longer than usual.
2 to remain alive or persist.}
&
\textgreater 10000 \\
\midrule

\textbf{Centroid 3} &
There is a reason why it makes sense to use 12 am for midnight: because it’s before noon of the same day. 
This reinforces the idea that the stroke of midnight begins the new day. 
As for 12 pm, it’s true that it’s at noon, not after it, but wait!
\textcolor{red}{imposed
imposed refers to something unwelcome or unpleasant that must be endured.}
&
\textgreater 10000 \\
\midrule

\multicolumn{3}{l}{\textbf{Stage 2: Optimized Adversarial Document (word-level substitutions highlighted in green)}} \\
\midrule

\textbf{Final Adversarial Document} &
There is a reason why it makes sense to use 12 am for midnight: because it’s before noon of the same day. 
This reinforces the idea that the \textcolor{darkgreen}{stroking} of midnight begins the new day. 
As for 12 pm, it’s true that it’s at noon, not after it, but wait!
\textcolor{red}{linger -- definition and \textcolor{darkgreen}{nouns}
This is the British English definition of linger.
View American English definition of linger.}
Change your default \textcolor{darkgreen}{vocabulary} to American English.
View the \textcolor{darkgreen}{vowels} for linger.
&
1 \\
\bottomrule
\end{tabular}

\caption{\textbf{Hard-level case study for the query ``hillsborough community college president.''} 
The target document is initially extremely irrelevant (not ranked in top-10000). 
After concatenation with centroid documents, the initial candidate rankings vary (243, outside top-10000, outside top-10000), 
and the best candidate is further refined through synonym substitutions, 
resulting in a final ranking of 1.}
\label{app:tab:hard_case}
\end{table*}

\clearpage


\end{document}